\documentclass[showpacs,aps,prd,nofootinbib,floatfix,amsmath,amssymb,twocolumn]{revtex4}
\usepackage{mathrsfs}
\usepackage{graphicx}
\usepackage[dvipsnames]{xcolor}
\usepackage{dsfont}
\begin{document}

\makeatletter
\newbox\slashbox \setbox\slashbox=\hbox{$/$}
\newbox\Slashbox \setbox\Slashbox=\hbox{\large$/$}
\def\pFMslash#1{\setbox\@tempboxa=\hbox{$#1$}
  \@tempdima=0.5\wd\slashbox \advance\@tempdima 0.5\wd\@tempboxa
  \copy\slashbox \kern-\@tempdima \box\@tempboxa}
\def\pFMSlash#1{\setbox\@tempboxa=\hbox{$#1$}
  \@tempdima=0.5\wd\Slashbox \advance\@tempdima 0.5\wd\@tempboxa
  \copy\Slashbox \kern-\@tempdima \box\@tempboxa}
\def\FMslash{\protect\pFMslash}
\def\FMSlash{\protect\pFMSlash}
\def\miss#1{\ifmmode{/\mkern-11mu #1}\else{${/\mkern-11mu #1}$}\fi}
\makeatother

\title{Majorana neutrinos in the triple gauge boson coupling $ZZZ^*$}

\author{H\'ector Novales-S\'anchez$^{(a)}$}
\author{M\'onica Salinas$^{(b)}$}
\affiliation{
$^{(a)}$Facultad de Ciencias F\'isico Matem\'aticas, Benem\'erita Universidad Aut\'onoma de Puebla, Apartado Postal 1152 Puebla, Puebla, M\'exico\\$^{(b)}$Departamento de F\'isica, Centro de Investigaci\'on y de Estudios Avanzados del IPN, Apartado Postal 14-740,07000 Ciudad de M\'exico, M\'exico}

\begin{abstract}
Confirmed by the measurement of neutrino oscillations, neutrino mass is recognized as a genuine manifestation of physics beyond the Standard Model, while its originating mechanism remains a mystery. Moreover, the proper field-theory description of neutrinos, whether they are Majorana or Dirac type, must be linked to such a mechanism. The present work addresses the calculation, estimation, and analysis of one-loop contributions from virtual Majorana neutrinos, light and heavy as well, to the neutral gauge boson coupling $ZZZ$, which participates in $Z$-boson pair production from $e^+e^-$ collisions. This task is carried out in the framework defined by a seesaw variant in which light neutrinos remain massless at tree level, then becoming massive radiatively. The $ZZZ^*$ coupling, with $Z^*$ an off-shell $Z$ boson, is defined by two form factors, namely, $f_4$, characterizing CP-odd effects, and  $f_5$, which is CP-even. Constraints from the Large Hadron Collider on both these quantities are currently ${\cal O}(10^{-4})$. Our calculation yields CP-nonpreserving contributions to $ZZZ$, which are absent in the framework of the sole Standard Model. Our estimations show that the $f_4$ contribution might be as large as ${\cal O}(10^{-7})$ for heavy-neutrino masses $\sim1\,{\rm TeV}$. CP-even contributions $f_5$ are also generated, which are, in general, larger than their CP-odd counterparts. We estimate them to be as large as ${\cal O}(10^{-4})$ at a center-of-mass energy of $500\,{\rm GeV}$, in $e^+e^-$ collisions. 
\end{abstract}


\maketitle

\section{Introduction}
\label{intro}
So far, the Standard Model~\cite{SMGlashow,SMSalam,SMWeinberg} (SM) is in good agreement with most experimental data~\cite{PDG}. In fact, the measurement~\cite{ATLASHiggs,CMSHiggs}, by the ATLAS and CMS experiments, of its last missing piece, the Higgs boson, has reinforced our trust in this formulation. Despite the broad success of the SM, this physical description is not, by any means, the last word, as experimentally-supported phenomena which are not properly explained by the SM exist, among which neutrino mass and mixing, dark matter, and dark energy are noteworthy. Thus, the intense work on the theoretical, phenomenological, and experimental fronts, aimed at the identification and estimation of possible manifestations of new physics, is well motivated, for it may provide us with hints about the genuine underlying high-energy formulation. 
In this context, the pursuit of such fundamental physical description, presumably governing nature at some high-energy scale, often relies in the exploration of observables which are suppressed, or even forbidden, in the framework established by the SM.
\\

The definition of the SM neutrino sector includes the assumption that neutrinos are massless, which in several situations works fine as an approximation, but fails as a correct description. The phenomenon of neutrino oscillations has been interpreted as a proof that neutrinos are massive and mix~\cite{Pontecorvo}. The observation of neutrino oscillations, first achieved by the Kamiokande Collaboration~\cite{KamiokandeNobel} and shortly after confirmed by the SNO Collaboration~\cite{SNONobel}, played a crucial role in solving the solar-neutrino problem. Since neutrino oscillations require neutrino mixing to happen, experimental efforts were devoted to determine the corresponding mixing angles, which concluded with the measurement of the last of such angles, $\theta_{13}$, by the Daya Bay Collaboration~\cite{DayaBaynuoscillations} and by the RENO Collaboration~\cite{RENOnuoscillations}. Now that neutrinos are known to have nonzero masses, a natural next step would be the determination of the origin of neutrino mass. Moreover, since neutrinos are electrically neutral and massive, their description corresponds to either Dirac fermions~\cite{DiracNobel} or Majorana fields~\cite{Majorana}. Among the whole set of known elementary fermions, neutrinos have, by far, the smallest masses, recently upper-bounded by the KATRIN Collaboration to be $\lesssim0.8\,{\rm eV}$~\cite{KATRINnumass}, which should be taken into account by any beyond-Standard-Model (BSM) proposal aiming at a sensible description of neutrinos. A nice and elegant explanation of neutrino mass is given by the seesaw mechanism~\cite{MoSe1,MoSe2}, with neutrinos characterized by Majorana fields. From the viewpoint of effective theories, the occurrence of this mechanism is underpinned by the Weinberg operator~\cite{Weinbergoperator}. In the context of the seesaw mechanism, besides the known neutrinos, a set of heavy partners, which we refer to as heavy neutrinos, arises. It turns out that the masses of these heavy neutrinos are restricted to be $\sim10^{13}\,{\rm GeV}$, in order for the seesaw mechanism to naturally generate tiny masses for the known light neutrinos. So, while the seesaw mechanism offers a pleasing explanation for the generation of neutrino mass, the presence of huge heavy-neutrino masses largely suppresses contributions, thus pushing presumed new-physics signals well beyond current experimental sensitivity. As a response, variants of the seesaw mechanism, such as the inverse seesaw mechanism~\cite{MoVa,GoVa,DeVa}, have emerged, in which the masses of heavy neutrinos are not so large, thus enhancing the effects of new physics. Another seesaw-mechanism variant was given by the author of Ref.~\cite{Pilaftsis}, who established a condition under which light-neutrino masses vanish at the tree level, thus weakening the link connecting light and heavy masses of neutrinos. Then masses of light neutrinos are properly defined at the loop level, as long as the spectrum of heavy-neutrino masses is quasi-degenerate. The theoretical set up of Ref.~\cite{Pilaftsis} is the framework of the present investigation.
\\

Triple gauge couplings (TGCs) are well-established places to look for traces of new physics. Among them, the SM $W$ boson electromagnetic and weak interactions $WW\gamma$ and $WWZ$, characterized by well-known general Lorentz-covariant parametrizations of their corresponding vertex functions~\cite{HPZH,BaZe}, have been widely studied in the SM and in several of its extensions as well. On the other hand, the gauge couplings $ZZZ$, $ZZ\gamma$, and $Z\gamma\gamma$, associated to electromagnetic and weak properties of neutral gauge bosons\footnote{Furry's theorem~\cite{Furry} forbids the occurrence of $\gamma\gamma\gamma$ as long as Lorentz symmetry holds~\cite{KLP}.}, bear their own appeal. The gauge structure of the SM precludes these neutral-gauge-boson interactions from happening at the tree level, in contraposition with the charged TGCs $WW\gamma$ and $WWZ$. Moreover, as a consequence of Bose symmetry (BS), the neutral TGCs vanish at the loop level whenever all the external particles are assumed to be on the mass shell, so, in order to generate nonzero contributions from the SM or from BSM physics, the calculation of any of these interactions must be executed by taking at least one of the external neutral bosons off the mass shell. Lorentz-covariant parametrizations of neutral TGCs, with the proper implementation of electromagnetic gauge symmetry and BS, were first given in Ref.~\cite{HPZH}, while these parametrizations were afterward readdressed in Refs.~\cite{GLR,BFFRS}. The one-loop contributions from the SM to the neutral TGCs were calculated and estimated in Refs.~\cite{GLR,CDRR}, finding that such contributions range from $\sim10^{-4}$ to $\sim10^{-3}$. Contributions generated by SM extensions can be found in the literature~\cite{GLR,CDRR,LPTT,CKP,DGM,Degrande,MTT}. Within the theoretical framework defined by the neutrino model of Ref.~\cite{Pilaftsis}, the present investigation considers the contributions, at one loop, from Majorana neutrinos, both light and heavy, to neutral TGCs. Since neutrinos do not couple to the electromagnetic field at the tree level, only contributions to $ZZZ$ are generated. Furthermore, our calculations, estimations and analyses are executed by thinking of the $ZZZ$ coupling as part of an $s$-channel diagram contributing to $Z$-boson pair production from an electron-positron collision. Therefore, the neutral TGC to calculate is $ZZZ^*$, with $Z^*$ denoting an off-shell $Z$ boson, in which case the general parametrization for this coupling is determined by two form factors, one of which is CP even whereas the other is CP odd. All the contributing Feynman diagrams are made of virtual-neutrino closed loops in which both light and heavy neutrinos participate. Despite the superficial degree of divergence of these diagrams, from the onset indicating the presumable presence of ultraviolet (UV) divergences, the contributions turn out to be finite. As opposed to the SM, these virtual-neutrino contributions are able to produce a nonzero CP-violating form factor, which is partly a consequence of the presence of couplings $Zn_jn_k$ where the neutrino fields $n_j$ and $n_k$ do not coincide. According to our estimations, contributions to the CP-odd form factor might be as large as $\sim10^{-7}$, for heavy-neutrino masses $\approx1.2\,{\rm TeV}$ and $\approx1.4\,{\rm TeV}$. Physical processes derived from electron-positron collisions, to take place in future $e^+e^-$ colliding machines, have been discussed by taking the value of $500\,{\rm GeV}$ for the CME as a reference~\cite{ILCandLHC,ILCtechnicalreport,BSZ,RaSi}. In particular, Ref.~\cite{RaSi} provides estimations of the sensitivity of a future electron-positron collider to neutral TGCs. A CP-conserving contribution is also generated from this neutrino model, which we find to be, at a CME of $500\,{\rm GeV}$, as large as $\sim10^{-4}$. For the sake of comparison, notice that the CMS Collaboration has recently bounded such quantities to lie below $\sim10^{-4}$~\cite{CMSboundsZZZ}. 
\\

The paper has been organized as follows: in Section~\ref{model}, the theoretical set up, as defined in Ref.~\cite{Pilaftsis}, is discussed; the Lorentz-covariant parametrization of $ZZZ^*$, as well as the analytic contributions from the neutrino model under consideration, are addressed in Section~\ref{phenoanalytic}; then, Section~\ref{phenonumeric} is devoted to numerical estimations and the analysis of the $ZZZ^*$ contributions; in the final part of the paper, Section~\ref{concs}, with give our conclusions and a summary.
\\

\section{Majorana neutrinos in BSM physics}
\label{model}
For several years, since their introduction in 1930, it was not certain whether neutrinos were massive or not, as the electric neutrality that characterizes these particles hindered a measurement of neutrino masses. Indeed, the neutrino fields in the SM were assumed to be chiral and massless, since the Weyl theory of massless fermions is consistent with non-observation of right-handed neutrino states. It was pointed out, in Ref.~\cite{Majorana}, that fermions which were both massive and neutral might abide by the Majorana condition, meaning that the involved fermion field, $\psi$, coincides with its corresponding charge-conjugate field, $\psi^{\rm c}=C\overline{\psi}^{\rm T}$, where $C$ is the charge-conjugation matrix. Later, in the 1950's, the possibility that neutrinos oscillate was posed~\cite{Pontecorvo}, which required nonzero neutrino masses and the occurrence of neutrino mixing. The first experimental evidence of neutrino oscillations was reported by the Kamiokande Collaboration~\cite{KamiokandeNobel}, in 1998, which would be corroborated by experiments at the Sudbury Neutrino Observatory~\cite{SNONobel}, in 2002. Regarding which description among Dirac and Majorana is the one faithfully characterizing neutrinos, there is not an answer yet. A number of experimental facilities have been pursuing the elusive neutrinoless double beta decay~\cite{nodoublebeta,CUPIDMo0nu2beta,CUORE0nu2beta,GERDA0nu2beta,MAJORANA0nu2beta,EXO0nu2beta,KamLANDZen0nu2beta}, which requires the Majorana description to happen in order to avoid final state neutrinos in this process. In this sense, a measurement of the neutrinoless double beta decay would be considered as definitive evidence in favor of the Majorana neutrino description. However, the large amount of experimental work aiming at the observation of this physical process has not succeeded so far~\cite{nodoublebeta}, and meanwhile the restrictive lower bound $10^{26}\,{\rm yr}$, on the neutrinoless double-beta decay half life, has been established by the GERDA Collaboration~\cite{GERDA0nu2beta} and by the KamLAND-Zen Collaboration~\cite{KamLANDZen0nu2beta}. 
\\

The so-called minimally extended SM~\cite{GiKi}, which yields the simplest approach to neutrino-mass generation,  gives rise to neutrino masses just as the SM does for the rest of the fermions, that is, through the introduction of right-handed Dirac-neutrino chiral fields and Yukawa neutrino terms affected by the Brout-Englert-Higgs mechanism~\cite{EnBr,Higgs}, with the values of the masses determined by both the electroweak scale, $v=246\,{\rm GeV}$, and a set of Yukawa constants. Nevertheless, a quite noticeable feature of neutrinos, which distinguishes them from every other known fermion, is the conspicuous smallness of their masses, currently upper-bounded to be within the sub-eV scale~\cite{KATRINnumass}. Such a characteristic has motivated the search for a more natural and reasonable explanation for the origin of neutrino mass. The Weinberg operator~\cite{Weinbergoperator}, ${\cal L}_{\rm W}=-\frac{\kappa_{\alpha\beta}}{\Lambda}\big( \overline{L^{\rm c}_{\alpha,L}}\tilde{\phi}^* \big)\big( \tilde{\phi}^\dag L_{\beta,L} \big)+{\rm H.c.}$, is an effective-Lagrangian term with units $({\rm mass})^5$, which is allowed only as long as lepton-number symmetry is violated. In this equation, $L_{\alpha,L}$ is the $\alpha$-th ${\rm SU}(2)_L$ lepton doublet, with left chirality, of the SM, whereas $\phi$ is the Higgs doublet, where $\tilde{\phi}=i\sigma^2\phi^*$, with $\sigma^2$ the imaginary Pauli matrix. Moreover, $\Lambda$ is interpreted as a high-energy scale, characterizing some BSM fundamental description, and $\kappa_{\alpha\beta}$ are dimensionless coefficients parametrizing the effects of such a high-energy formulation at the level of low energies. After electroweak symmetry breaking, the Weinberg operator engenders Majorana mass terms for neutrinos, with masses suppressed by the energy scale $\Lambda$. This suggests that some high-energy scale, at which the BSM fundamental formulation operates in full, would be responsible for the tininess of neutrino masses. Left-right symmetric models, based on the gauge group ${\rm SU}(3)_C\otimes{\rm SU}(2)_L\otimes{\rm SU}(2)_R\otimes{\rm U}(1)_{B-L}$, are SM extensions originally meant to address parity violation in low-energy processes~\cite{PaSa,MoPa1,MoPa2}. A notorious feature of left-right models turned out to be the seesaw mechanism~\cite{MoSe1,MoSe2}, by which, after a couple of stages of spontaneous symmetry breaking, masses can be defined for Majorana neutrino fields. In general, besides the three known neutrinos, the seesaw mechanism gives rise to a further set of neutrinos. The masses of known neutrinos, given by the seesaw mechanism, are found to follow the Weinberg-operator profile, since they are given as $m_\nu\sim \frac{v^2}{w}$, where $w$ is a high-energy scale, in this case the one at which some high-energy phase of spontaneous symmetry breaking takes place. It is worth emphasizing the suppression induced by the scale $w$ on these masses. The masses of the new neutrinos, on the other hand, differ dramatically in the sense that they are not diminished by the high-energy scale $w$, but they are rather proportional to it, that is $m_N\sim w$. Thus, the larger the masses of new neutrinos, the smaller the masses of known neutrinos. Keeping this in mind, in what follows we use the terms ``light neutrinos'' and ``heavy neutrinos'' to distinguish known neutrinos from new neutrinos, respectively. 
\\

While nonzero masses of light neutrinos are nicely explained by the seesaw mechanism, current upper bounds on such masses impose very strict constraints on the high-energy scale $w$, which goes up to $\sim10^{13}\,{\rm GeV}$. An energy scale so large renders heavy-neutrino masses, $m_N\sim w$, huge, thus severely attenuating the impact of these particles on physical processes attainable by current experimental facilities, then leaving the possibility of measuring their effects in the near future off the table. Aiming at bettering scenarios of neutrino mass generation in the presence of heavy neutrinos, variations of the seesaw mechanism have been conceived and explored. In the inverse seesaw~\cite{MoVa,GoVa,DeVa}, for instance, besides three heavy neutrinos, another set of Majorana neutral fermions is introduced, which, together with assumptions on the couplings of the neutrino fields, leads to a non-diagonal mass matrix whose structure matches the one corresponding to type-1 seesaw. Block matrices nested within such a mass matrix provide parameters, assumed to be small, which weakens the link between the high-energy scale $w$ and the neutrino masses. This framework turns out to be appealing, as more reasonable heavy-neutrino masses become allowed. There are further seesaw variations from which flexible values of heavy-neutrino masses can be defined, as it is the case of the model given in Ref.~\cite{Pilaftsis}. The investigation discussed throughout the present paper has been carried out within the framework of this reference, which we discuss below.
\\

Think of a BSM high-energy physical description, distinguished by the Lagrangian density ${\cal L}_{\rm BSM}$. Such a model might be governed by an extended gauge-symmetry group, as it is the case of left-right models~\cite{MoSe1,MoSe2,MoPa1,MoPa2}, 331 models~\cite{PiPl,Frampton}, and grand unification models~\cite{GeGl,GQW}. Let us assume that this formulation undergoes two stages of spontaneous symmetry breaking to finally fall into the electromagnetic group, ${\rm U}(1)_e$, characterized by electromagnetic gauge invariance. Imagine that the first phase of symmetry breaking, taking place at $w$, renders ${\cal L}_{\rm BSM}$ invariant with respect to the SM gauge symmetry group, ${\rm SU}(3)_C\otimes{\rm SU}(2)_\otimes{\rm U}(1)_Y$. Then, at $v$, the Englert-Brout-Higgs mechanism operates, thus yielding a Lagrangian 
\begin{equation}
{\cal L}_{\rm BSM}={\cal L}_{\rm mass}^\nu+{\cal L}^W_{\rm CC}+{\cal L}^Z_{\rm NC}+\ldots
\label{LBSMafter}
\end{equation}
The ellipsis in Eq.~(\ref{LBSMafter}) represents a set of Lagrangian terms, which are, in general, distinctive of the SM extension under consideration, as they may depend on non-SM dynamic variables or involve couplings dictated by the symmetries of ${\cal L}_{\rm BSM}$ at high energies, before the occurrence of any stage of spontaneous symmetry breaking. The Lagrangian term ${\cal L}_{\rm mass}^\nu$, which gathers all the couplings of the theory that are quadratic in neutrino fields, is assumed to be given by
\begin{eqnarray}
{\cal L}_{\rm mass}^\nu=&
\displaystyle
-\sum_{j=1}^3\sum_{k=1}^3\Big( \overline{\nu^0_{j,L}} (m_{\rm D})_{jk}\,\nu^0_{k,R}
\nonumber \\
&
\displaystyle
+\frac{1}{2}\overline{\nu_{j,R}^{0{\rm c}}}(m_{\rm M})_{jk}\,\nu^0_{k,R} \Big)+{\rm H.c.}
\label{Lnumass}
\end{eqnarray}
This equation involves three left-handed neutrino fields, $\nu^0_{j,L}$, as well as three right-handed neutrino fields, $\nu^0_{j,R}$. The charge-conjugate fields $\nu_{j,R}^{0{\rm c}}=C\overline{\nu^0_{j,R}}^{\rm T}$, with left-handed chirality, also appear in Eq.~(\ref{Lnumass}), being part of a set of Majorana-like mass terms, which involve neutrino mixing. The mixing featured in these terms is characterized by the $3\times3$ matrix $m_{\rm M}$, which, due to the Majorana condition $\nu^{0{\rm c}}_{j,R}=\nu^0_{j,R}$, is symmetric, while note that this matrix is general in any other respect. The Majorana matrix $m_{\rm M}$ is assumed to emerge as a consequence of the first symmetry breaking, at $w$. On the other hand, Dirac-like mass terms, also included in ${\cal L}_{\rm mass}^\nu$, are given by neutrino-field mixing through the Dirac-mass matrix $m_{\rm D}$, which is $3\times3$ sized and general. We assume the matrix $m_{\rm D}$ to originate from electroweak symmetry breaking. The column matrices
\begin{eqnarray}
f_L=
\left(
\begin{array}{c}
\nu^0_{1,L}
\vspace{0.2cm}
\\
\nu^0_{2,L}
\vspace{0.2cm}
\\
\nu^0_{3,L}
\end{array}
\right),
&&
F_L=
\left(
\begin{array}{c}
\nu^{0{\rm c}}_{1,R}
\vspace{0.2cm}
\\
\nu^{0{\rm c}}_{2,R}
\vspace{0.2cm}
\\
\nu^{0{\rm c}}_{3,R}
\end{array}
\right),
\\ \\
f_R=
\left(
\begin{array}{c}
\nu^{0,{\rm c}}_{1,L}
\vspace{0.2cm}
\\
\nu^{0,{\rm c}}_{2,L}
\vspace{0.2cm}
\\
\nu^{0,{\rm c}}_{3,L}
\end{array}
\right),
&&
F_R=
\left(
\begin{array}{c}
\nu^0_{1,R}
\vspace{0.2cm}
\\
\nu^0_{2,R}
\vspace{0.2cm}
\\
\nu^0_{3,R}
\end{array}
\right),
\end{eqnarray}
with $f_R=f_L^{\rm c}$ and $F_R=F_L^{\rm c}$, are defined and utilized to rearrange ${\cal L}_{\rm mass}^\nu$ as
\begin{equation}
{\cal L}_{\rm mass}^\nu=-\frac{1}{2}\Big( \overline{f_L}\hspace{0.2cm}\overline{F_L} \Big){\cal M}
\left(
\begin{array}{c}
f_R
\vspace{0.2cm}
\\
F_R
\end{array}
\right)
+{\rm H.c.}
\end{equation}
Here, ${\cal M}$ is a $6\times6$ matrix, which is conveniently written in block-matrix form as
\begin{equation}
{\cal M}=
\left(
\begin{array}{cc}
0 & m_{\rm D}
\vspace{0.3cm}
\\
m_{\rm D}^{\rm T} & m_{\rm M}
\end{array}
\right).
\label{Mbefdiag}
\end{equation}
The structure of the non-diagonal mass matrix ${\cal M}$, displayed in the last equation, corresponds to type-1 seesaw mechanism. Since $m_{\rm M}^{\rm T}=m_{\rm M}$ holds, the matrix ${\cal M}$ turns out to be symmetric, which implies that a diagonalization $6\times6$ unitary matrix, ${\cal U}_\nu$, yielding
\begin{equation}
{\cal U}_\nu^{\rm T}{\cal M}\,{\cal U}_\nu=
\left(
\begin{array}{cc}
m_\nu & 0
\vspace{0.3cm}
\\
0 & m_N
\end{array}
\right),
\label{UMU}
\end{equation}
exists~\cite{Takagi}, where $m_\nu$ and $m_N$ are diagonal and real $3\times3$ matrices, with $m_{\nu_j}=(m_\nu)_{jj}>0$ and $m_{N_j}=(m_N)_{jj}>0$, for $j=1,2,3$. By means of this diagonalization, the neutrino mass-eigenfields basis $\{ \nu_1,\nu_2,\nu_3,N_1,N_2,N_3 \}$ is defined, in terms of which the neutrino-mass Lagrangian adopts the form
\begin{equation}
{\cal L}_{\rm mass}^\nu=\sum_{j=1}^3
\Big(
-\frac{1}{2}m_{\nu_j}\overline{\nu_j}\,\nu_j-\frac{1}{2}m_{N_j}\overline{N_j}\,N_j
\Big).
\end{equation}
Note that these neutrino fields are Majorana spinors, since they fulfill the Majorana condition, $\nu_j^{\rm c}=\nu_j$ and $N_j^{\rm c}=N_j$.
\\

Let us express the $6\times6$ diagonalization matrix ${\cal U}_\nu$ as a block matrix made of $3\times3$ matrix blocks, ${\cal U}_{jk}$, that is,
\begin{equation}
{\cal U}=
\left(
\begin{array}{cc}
{\cal U}_{11} & {\cal U}_{12}
\vspace{0.3cm}
\\
{\cal U}_{21} & {\cal U}_{22}
\end{array}
\right).
\end{equation}
Next, the following quantities are defined:
\begin{eqnarray}
{\cal B}_{\alpha\nu_j}=\sum_{k=1}^3V^\ell_{\alpha k}({\cal U}^*_{11})_{kj},
\label{Banu}
\\
{\cal B}_{\alpha N_j}=\sum_{k=1}^3V^\ell_{\alpha k}({\cal U}^*_{12})_{kj}.
\label{BaN}
\end{eqnarray}
In these equations, the greek index $\alpha$ labels SM lepton flavors, thus meaning that $\alpha=e,\mu,\tau$. On the other hand, $V^\ell$ is a $3\times3$ matrix, which is a lepton-sector analogue of the SM Kobayashi-Maskawa quark-mixing matrix~\cite{KoMa}. It is worth keeping in mind, though, that $V^\ell$ is not necessarily unitary, which contrasts with the unitarity feature characterizing the Pontecorvo-Maki-Nakagawa-Sakata (PMNS) mixing matrix~\cite{MNS,PontecorvoPMNS}, ${\cal U}_{\rm PMNS}$. The PMNS matrix is used to handle lepton mixing when only the three light neutrinos participate. Usage of this matrix is suitable for SM extensions in which the presence of heavy neutrinos can be disregarded at low energies, which, for instance, is the case of the Weinberg operator. Eqs.~(\ref{Banu}) and (\ref{BaN}) define the matrices ${\cal B}_\nu$ and ${\cal B}_N$, both $3\times3$ sized, through their entries $({\cal B}_\nu)_{\alpha\nu_j}={\cal B}_{\alpha\nu_j}$ and $({\cal B}_N)_{\alpha N_j}={\cal B}_{\alpha N_j}$. These two matrices are gathered into the $3\times6$ matrix ${\cal B}=\big( {\cal B}_\nu\hspace{0.2cm}{\cal B}_N \big)$, whose entries read
\begin{equation}
{\cal B}_{\alpha j}=
\left\{
\begin{array}{l}
{\cal B}_{\alpha \nu_k},\,\,\textrm{if}\,\,j=1,2,3,
\vspace{0.4cm}
\\
{\cal B}_{\alpha N_k},\,\,\textrm{if}\,\,j=4,5,6,
\end{array}
\right.
\end{equation}
with $\nu_k=\nu_1,\nu_2,\nu_3$ and $N_k=N_1,N_2,N_3$. Moreover, the matrix ${\cal B}$ satisfies the conditions
\begin{eqnarray}
\sum_{k=1}^6{\cal B}_{\alpha k}{\cal B}^*_{\beta k}=\delta_{\alpha\beta},
\label{BBcond1}
\\
\sum_{\alpha=e,\mu,\tau}{\cal B}^*_{\alpha j}{\cal B}_{\alpha k}={\cal C}_{jk},
\label{BBcond2}
\end{eqnarray}
with $\delta_{\alpha\beta}=({\bf 1}_3)_{\alpha\beta}$, where ${\bf 1}_3$ is the $3\times3$ identity matrix. In these equations, ${\cal C}_{jk}$ are the entries of a a $6\times6$ matrix, ${\cal C}$. This matrix is conveniently written in block-matrix form as
\begin{equation}
{\cal C}=
\left(
\begin{array}{cc}
{\cal C}_{\nu\nu} & {\cal C}_{\nu N}
\vspace{0.4cm}
\\
{\cal C}_{N\nu} & {\cal C}_{NN}
\end{array}
\right)
\end{equation}
with the $3\times3$ matrix blocks given by
\begin{eqnarray}
&&
({\cal C}_{\nu\nu})_{il}\equiv{\cal C}_{\nu_i\nu_l}=\sum_{j=1}^3({\cal U}_{11})_{ji}({\cal U}_{11}^*)_{jl},
\\&&
({\cal C}_{\nu N})_{il}\equiv{\cal C}_{\nu_iN_l}=\sum_{j=1}^3({\cal U}_{11})_{ji}({\cal U}_{12}^*)_{jl},
\\&&
({\cal C}_{N\nu})_{il}\equiv{\cal C}_{N_i\nu_l}=\sum_{j=1}^3({\cal U}_{12})_{ji}({\cal U}_{11}^*)_{jl},
\\&&
({\cal C}_{NN})_{il}\equiv{\cal C}_{N_iN_l}=\sum_{j=1}^3({\cal U}_{12})_{ji}({\cal U}_{12}^*)_{jl}.
\end{eqnarray}
While Eq.~(\ref{BBcond1}) resembles a unitarity condition, notice that Eq.~(\ref{BBcond2}) shows that, strictly speaking, this is not the case. In matrix form, Eqs.~(\ref{BBcond1}) and (\ref{BBcond2}) are succinctly expressed as ${\cal B}{\cal B}^\dag={\bf 1}_3$ and ${\cal B}^\dag{\cal B}={\cal C}$, respectively. The matrix ${\cal C}$, on the other hand, fulfills
\begin{equation}
\sum_{i=1}^6{\cal C}_{ji}{\cal C}^*_{ki}={\cal C}_{jk},
\end{equation}
also written, in matrix form, as ${\cal C}{\cal C}^\dag={\cal C}$.
\\

Besides the neutrino-mass Lagrangian term ${\cal L}_{\rm mass}^\nu$, the only other terms explicitly shown by Eq.~(\ref{LBSMafter}) are ${\cal L}_{\rm CC}^W$ and ${\cal L}_{\rm NC}^Z$, which are given by
\begin{eqnarray}
&&
{\cal L}_{\rm CC}^W
\displaystyle
=\sum_\alpha\sum_{j=1}^3
\Big(
\frac{g}{\sqrt{2}}{\cal B}_{\alpha\nu_j}W^-_\rho\overline{l_\alpha}\gamma^\rho P_L\nu_j
\nonumber \\ &&
\hspace{0.9cm}
\displaystyle
+\frac{g}{\sqrt{2}}{\cal B}_{\alpha N_j}W^-_\rho\overline{l_\alpha}\gamma^\rho P_LN_j
\Big)
+{\rm H.c.},
\label{SMWCCs}
\end{eqnarray}
\begin{eqnarray}
&&
{\cal L}^Z_{\rm NC}=
\displaystyle
\sum_{k=1}^3\sum_{j=1}^3
\Big(
-\frac{g}{4c_{\rm W}}Z_\rho\,\overline{\nu_k}\gamma^\rho\big( i{\cal C}^{\rm Im}_{\nu_k\nu_j}-{\cal C}^{\rm Re}_{\nu_k\nu_j}\gamma_5 \big)\nu_j
\nonumber \\ && 
\hspace{0.9cm}
\displaystyle
+
\Big(
-\frac{g}{4c_{\rm W}}Z_\rho\,\overline{\nu_k}\gamma^\rho\big( i{\cal C}^{\rm Im}_{\nu_kN_j}-{\cal C}^{\rm Re}_{\nu_kN_j}\gamma_5 \big)N_j+{\rm H.c.}
\Big)
\nonumber \\ && 
\hspace{0.9cm}
\displaystyle
-\frac{g}{4c_{\rm W}}Z_\rho\,\overline{N_k}\gamma^\rho\big( i{\cal C}^{\rm Im}_{N_kN_j}-{\cal C}^{\rm Re}_{N_kN_j}\gamma_5 \big)N_j
\Big).
\label{SMZNCs}
\end{eqnarray}
In these equations, $g$ is the ${\rm SU}(2)_L$ coupling constant, whereas $c_{\rm W}=\cos\theta_{\rm W}$ denotes the cosine of the weak mixing angle, $\theta_{\rm W}$. Furthermore, $W_\rho$ is the SM $W$-boson field and $Z_\rho$ is the $Z$-boson field, also of the SM. We have denoted ${\rm Re}\{ {\cal C} \}={\cal C}^{\rm Re}$ and ${\rm Im}\{ {\cal C} \}={\cal C}^{\rm Im}$, so that ${\cal C}={\cal C}^{\rm Re}+i\,{\cal C}^{\rm Im}$. Eq.~(\ref{SMWCCs}) comprises all the charged-current terms which feature the SM $W$ boson. This Lagrangian term was recently utilized, in Ref.~\cite{MMNS}, to calculate and estimate the contributions, at one loop, from Majorana neutrinos, both light and heavy ones, to the TGC $WW\gamma$, in the context of the neutrino-mass model posed by the author of Ref.~\cite{Pilaftsis}. Neutral currents (NC) involving the SM $Z$ boson, on the other hand, are given by the couplings constituting the Lagrangian ${\cal L}^Z_{\rm NC}$, displayed in Eq.~(\ref{SMZNCs}). 
\\

Recall the Majorana and Dirac mass matrices $m_{\rm M}$ and $m_{\rm D}$, which are part of the non-diagonal mass matrix ${\cal M}$, as shown in Eq.~(\ref{Mbefdiag}). If we assume a scenario in which these matrices behave as $m_{\rm M}\sim w$ and $m_{\rm D}\sim v$, with the condition $v\ll w$ fulfilled, we get the type-1 seesaw mechanism, which, as discussed before, bears the disadvantage of marginal impact of the new physics. Aiming at an amelioration of this issue, the author of Ref.~\cite{Pilaftsis} considered the set of conditions $({\cal M}{\cal U}_\nu)_{jk}=0$, for $j=1,2,3,4,5,6$, which eliminate the tree-level mass of the $k$-th light neutrino. Such a condition was then implemented to cancel all light-neutrino mass terms from ${\cal L}_{\rm mass}^\nu$, thus meaning that the $3\times3$ diagonal matrix $m_\nu$, given by Eq.~(\ref{UMU}), vanishes. Let us comment that this procedure to render light-neutrino masses zero does not alter the analytic expressions for the masses of heavy neutrinos. In other words, the $3\times3$ matrix $m_N$, given by Eq.~(\ref{UMU}), remains the unchanged. In this context, light neutrinos become massive by quantum effects. The author of Ref.~\cite{Pilaftsis} calculated such masses from self-energy Feynman diagrams, at one loop, and provided the corresponding analytic expressions. The eradication of light-neutrino masses at the tree level, and their definition through loop diagrams, attenuates the connection among light- and heavy-neutrino masses, thus allowing for quite smaller heavy masses $m_{N_k}$ to occur. Moreover, in this model, the tininess of light neutrinos is rather an implication of the occurrence of a quasi-degenerate spectrum of heavy-neutrino masses~\cite{Pilaftsis}. 
\\

Let us work, from here on, in the framework discussed in the previous paragraph. The non-diagonal, though symmetric, mass matrix ${\cal M}$ can be block-diagonalized by the unitary matrix~\cite{KPS,DePi}
\begin{equation}
{\cal U}_\nu=
\left(
\begin{array}{cc}
\big( {\bf 1}_3+\xi^*\xi^{\rm T} \big)^{-\frac{1}{2}} & \xi^*\big( {\bf 1}_3+\xi^{\rm T}\xi^* \big)^{-\frac{1}{2}}
\vspace{0.4cm}
\\
-\xi^{\rm T}\big( {\bf 1}_3+\xi^*\xi^{\rm T} \big)^{\rm -\frac{1}{2}} & \big( {\bf 1}_3+\xi^{\rm T}\xi^* \big)^{-\frac{1}{2}}
\end{array}
\right),
\end{equation}
where $\xi$ is some $3\times3$ matrix. Assuming the moduli $|\xi_{jk}|$ to be small, the relation
\begin{equation}
\xi=m_{\rm D}m_{\rm M}^{-1},
\label{xidef}
\end{equation}
distinctive of the ordinary seesaw mechanism, holds. Furthermore, in this context, the diagonalization matrix ${\cal U}_\nu$ can be approximated as~\cite{Pilaftsis}
\begin{equation}
{\cal U}_\nu\simeq
\left(
\begin{array}{cc}
{\bf 1}_3-\frac{1}{2}\xi^*\xi^{\rm T} & \xi^*\big( {\bf 1}_3-\frac{1}{2}\xi^{\rm T}\xi^* \big)
\vspace{0.4cm}
\\
-\xi^{\rm T}\big( {\bf 1}_3-\frac{1}{2}\xi^*\xi^{\rm T} \big) & {\bf 1}_3-\frac{1}{2}\xi^{\rm T}\xi^*
\end{array}
\right),
\end{equation}
at ${\cal O}(\xi^3)$. Heavy-neutrino masses turn out to be~\cite{Pilaftsis}
\begin{equation}
m_N\simeq m_{\rm M}\Big( {\bf 1}_3+\frac{1}{2}m_{\rm M}^{-1}\big( \xi^\dag m_{\rm D}+m_{\rm D}^{\rm T}\xi^* \big) \Big).
\end{equation}
Furthermore, the matrix ${\cal B}$ is given by
\begin{equation}
{\cal B}\simeq\bigg( V^\ell\Big( {\bf 1}_3-\frac{1}{2}\xi\xi^\dag \Big)\hspace{0.4cm}V^\ell\xi\Big( {\bf 1}_3-\frac{1}{2}\xi^\dag\xi \Big) \bigg),
\end{equation}
whereas ${\cal C}$ acquires the form
\begin{equation}
{\cal C}\simeq
\left(
\begin{array}{ccc}
{\bf 1}_3-\xi\xi^\dag && \xi\big( {\bf 1}_3-\xi^\dag\xi \big)
\vspace{0.4cm}
\\
\Big( \xi\big( {\bf 1}_3-\xi^\dag\xi \big) \Big)^\dag && \xi^\dag\xi
\end{array}
\right).
\label{Cinxi}
\end{equation}
\\

\section{Contributions from Majorana neutrinos to $ZZZ^*$ at one loop}
\label{phenoanalytic}
In this Section, we carry out a calculation of the contributions from the neutrino model given in Ref.~\cite{Pilaftsis} to the vertex $ZZZ^*$, which points towards an estimation of the effects produced by virtual neutrino fields, both light and heavy, to the TGCs characterizing this interaction. In general, TGCs also emerge from neutral gauge bosons interactions in which external photon fields participate, as, for instance, is the case of $ZZA^*$. Nevertheless, notice that, in general, virtual-neutrino contributions to such couplings do not exist at the one-loop level, due to electric neutrality of neutrinos, while nonzero contributions at higher loop orders might emerge. Calculations beyond the one-loop level are not within the scope of the present work, so we concentrate in the virtual-neutrino contributions to $ZZZ^*$.

\subsection{The vertex $ZZZ^*$}
Consider the effective Lagrangian~\cite{GLRotro}
\begin{eqnarray}
&&
{\cal L}_{\rm eff}^{ZZZ}=\frac{e}{m_Z^2}
\Big(
-f_4(\partial_\mu Z^{\mu\beta})Z_\alpha(\partial^\alpha Z_\beta)
\nonumber \\ && \hspace{1.5cm}
+f_5(\partial^\alpha Z_{\alpha\mu})\tilde{Z}^{\mu\beta}Z_\beta
\Big),
\end{eqnarray}
whose mass-dimension is 6. Then notice that the factor $(m_Z^2)^{-1}$, which is part of its definition, is intended to corrects units, thus meaning that the form factors $f_4$ and $f_5$ are dimensionless. Moreover, the 2-tensor $Z_{\mu\nu}=\partial_\mu Z_\nu-\partial_\nu Z_\mu$ is defined, with $\tilde{Z}_{\mu\nu}=\frac{1}{2}\epsilon_{\mu\nu\rho\sigma}Z^{\rho\sigma}$ its corresponding dual tensor. The $f_4$ Lagrangian term violates CP symmetry, whereas the $f_5$ term preserves it. Assume that the $ZZZ^*$ vertex is part of an $s$-channel diagram contributing to $Z$-boson pair production through a positron-electron collision. Recall that the symbol $Z^*$ indicates that this $Z$ boson is off the mass shell. In order to derive the vertex function corresponding to ${\cal L}_{\rm eff}^{ZZZ}$, we follow the conventions
\begin{equation}
ie\,\Gamma^{ZZZ^*}_{\alpha\beta\mu}=
\begin{gathered}
\vspace{-0.35cm}
\includegraphics[width=4cm]{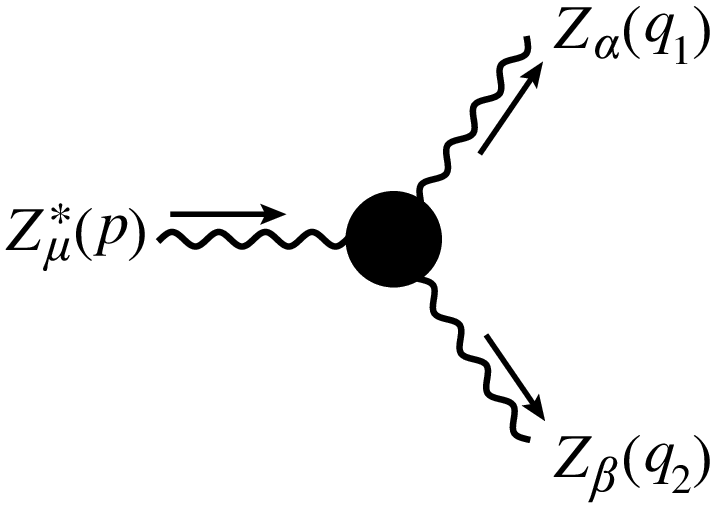}
\end{gathered}
\end{equation}
where the $ZZZ^*$ vertex is displayed. Keep in mind that the determination of the vertex function is carried out under the assumption that two $Z$ bosons are on shell, whereas the third one, with momentum $p=q_1+q_2$, is off shell. So, while $q_1^2=m_Z^2$ and $q_2^2=m_Z^2$, we denote $p^2=s=(q_1+q_2)^2$. BS must be taken into account for the vertex function to be correctly determined. The vertex function is given, under such circumstances, by
\begin{eqnarray}
&&
\Gamma^{ZZZ^*}_{\alpha\beta\mu}=\frac{if_4}{m_Z^2}
\Big(
(s-m_Z^2)\big( p_\alpha\,g_{\beta\mu}+p_\beta\,g_{\alpha\mu} \big)
\nonumber \\ && \hspace{1.3cm}
+p_\mu \big( m_Z^2\,g_{\alpha\beta}-2p_\alpha \,p_\beta \big)
\Big)
\nonumber \\ && \hspace{1.3cm}
-\frac{if_5}{m_Z^2}
\Big(
(s-m_Z^2)\epsilon_{\mu\alpha\beta\rho} (q_1^\rho-q_2^\rho)
\nonumber \\ && \hspace{1.3cm}
-p_\mu\epsilon_{\alpha\beta\lambda\rho}\,p^\lambda p^\rho
\Big).
\label{ZZZvfcomplete}
\end{eqnarray}
Note that any term in Eq.~(\ref{ZZZvfcomplete}) involve either the factor $s-m_Z^2$ or the 4-momentum component $p_\mu$, which in turn implies that the whole contribution vanishes if the three external $Z$ bosons are taken on shell, as both $s=m_Z^2$ and the transversality condition $p_\mu\epsilon^\mu(p)=0$, with $\epsilon^\mu(p)$ the polarization 4-vector, hold in such a context. Let us assume that this vertex connects with a conserved current $j^\mu$, so that $p_\mu j^\mu=0$ is valid as long as initial-state electron-positron masses are neglected. This assumption is customarily used~\cite{BFFRS}. This then leaves us with
\begin{eqnarray}
&&
\Gamma^{ZZZ^*}_{\alpha\beta\mu}=\frac{i(s-m_Z^2)}{m_Z^2}
\Big(
f_4\,\big( p_\alpha\,g_{\beta\mu}+p_\beta\,g_{\alpha\mu} \big)
\nonumber \\ && \hspace{1.3cm}
-f_5\,\epsilon_{\mu\alpha\beta\rho}(q_1^\rho-q_2^\rho)
\Big).
\label{ZZZvfunction}
\end{eqnarray}
which is the well-known Lorentz-covariant parametrization of this vertex~\cite{HPZH,GLR}. As we commented before, the form factor $f_4$ quantifies CP-odd effects. CP-symmetry non-preservation bears great relevance, not only because it is interesting by itself, but also because this phenomenon is, according to Sakharov criteria~\cite{Sakharov}, a requirement for the observed baryon asymmetry to be explained. Since not enough CP violation is provided by the SM, the presence and exploration of sources of this phenomenon is quite appealing. 
\\

\subsection{One-loop analytical contributions}
Throughout this subsection, the analytical calculation of one-loop contributions to $ZZZ^*$ from the neutrino model discussed in Section~\ref{model} is performed. At one loop, all the contributing Feynman diagrams involve virtual neutrinos, which combine into closed fermion loops. The necessary Feynman rules to assemble the contributing $ZZZ^*$ diagrams follow from the NC Lagrangian term ${\cal L}_{\rm NC}^Z$, displayed in Eq.~(\ref{SMZNCs}). In what follows, neutrino fields, light ones and heavy ones as well, are generically denoted by $n_i$, where $n_1=\nu_1$, $n_2=\nu_2$, $n_3=\nu_3$, $n_4=N_1$, $n_5=N_2$, and $n_6=N_3$. We conveniently express any term of ${\cal L}_{\rm NC}^Z$ as
\begin{equation}
{\cal L}_{Zn_kn_j}=-i\,Z_\mu\overline{n_k}\,\Gamma^\mu_{kj}\,n_j,
\label{Znknj}
\end{equation}
so ${\cal L}_{\rm NC}^Z=\sum_{k=1}^6\sum_{j=1}^6{\cal L}_{Zn_kn_j}$. By looking at these equations, note that couplings of the $Z$ boson to neutrino pairs mix neutrino fields, both light and heavy, so vertices $Znn$ change neutrino type.
\\

\begin{figure}[ht]
\center
\includegraphics[width=4cm]{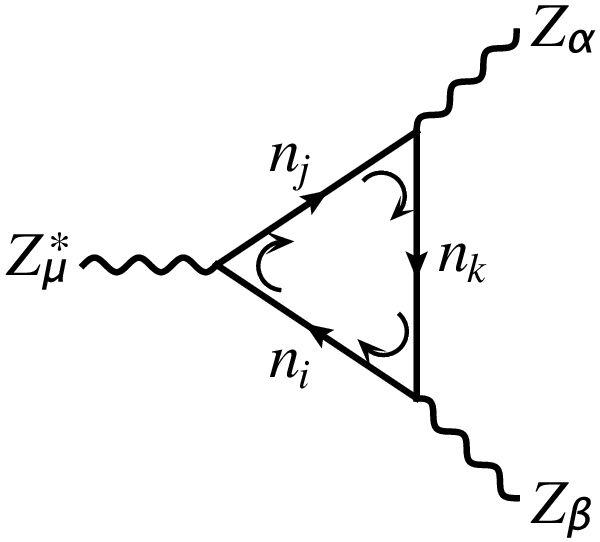}
\hspace{0.4cm}
\includegraphics[width=4cm]{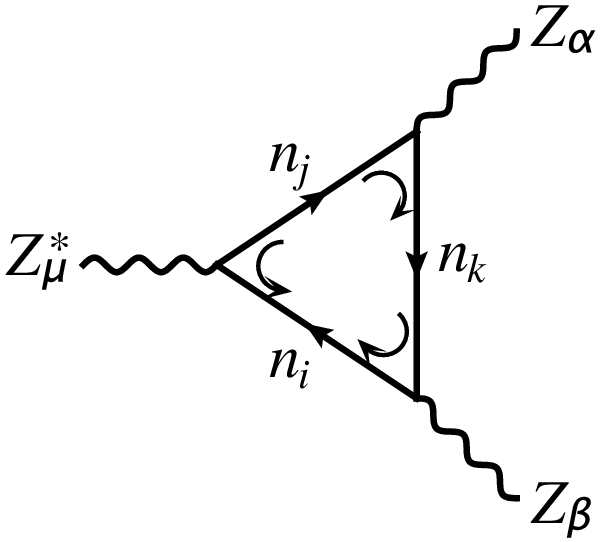}
\vspace{0.4cm} \\
\includegraphics[width=4cm]{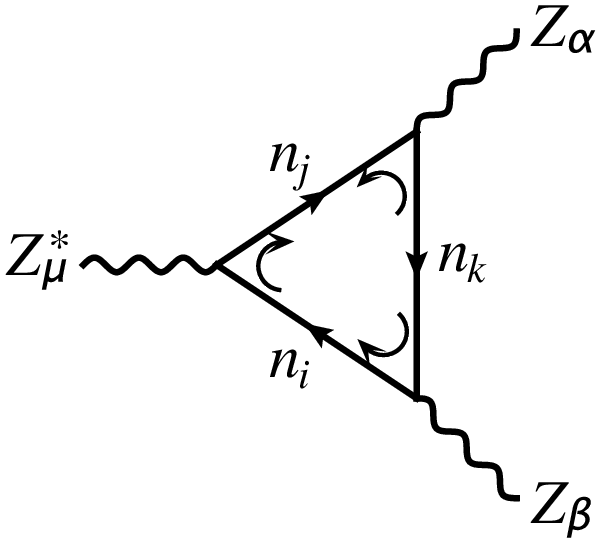}
\hspace{0.4cm}
\includegraphics[width=4cm]{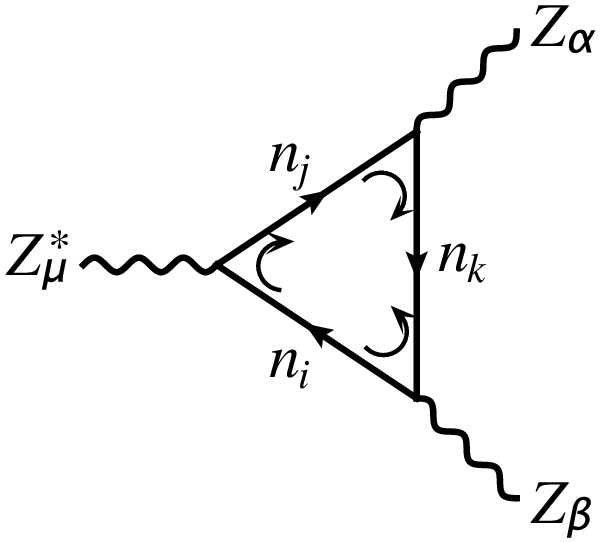}
\caption{\label{Feyndiags} A subset of all the Feynman diagrams contributing to $ZZZ^*$ at the one-loop level. The determination of all the missing diagrams is achieved by implementing BS to each of the diagrams shown in this figure.}
\end{figure}

The neutrino model under consideration comes along with the assumption that all the neutrinos are characterized by Majorana fields. Differences between the Dirac and Majorana descriptions manifest at the level of Feynman diagrams, as the Feynman rules used in these two scenarios differ from each other. A useful discussion on Feynman rules in the presence of Majorana fields is given in Ref.~\cite{DEHK}. In particular, the number of diagrams contributing to a given physical process or observable is usually larger if neutrinos are Majorana, in comparison with the Dirac treatment. Consider the Feynman diagrams shown in Fig.~\ref{Feyndiags},
which constitute a subset of the complete collection of contributing diagrams. To determine these contributing diagrams, use has been made of the Wick's theorem~\cite{Wick}. Even though fermion number is not preserved by Majorana neutrinos, arrows on fermion lines have been added, which represent a reference flux direction, as suggested in Ref.~\cite{DEHK}. Curved arrows lying within loops, off fermion lines, are used to denote the type of vertex $Znn$ which has to be used to write down the analytic expression of the diagram. These arrows point in either the same or in the opposite direction of the reference flux. If the directions coincide, the vertex $Znn$ is the one directly obtained from the Lagrangian, which, in accordance with Eq.~(\ref{Znknj}), is given as
\begin{equation}
\begin{gathered}
\vspace{-0.055cm}
\includegraphics[width=2.5cm]{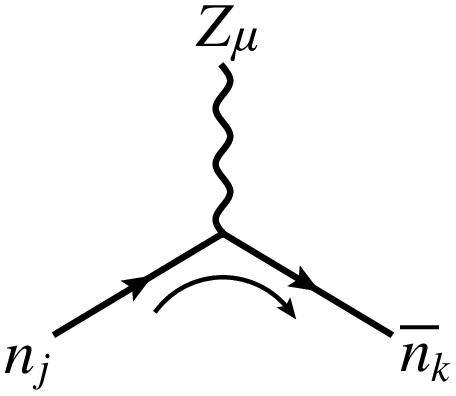}
\end{gathered}
=\Gamma^\mu_{kj}.
\end{equation}
According to Wick's theorem, among the curved arrows in the diagrams of Fig.~\ref{Feyndiags}, at most one can point oppositely to the reference neutrino flux. Vertices $Znn$ with a curved arrow pointing in the opposite direction of the reference fermion flux are given by
\begin{equation}
\begin{gathered}
\vspace{-0.055cm}
\includegraphics[width=2.5cm]{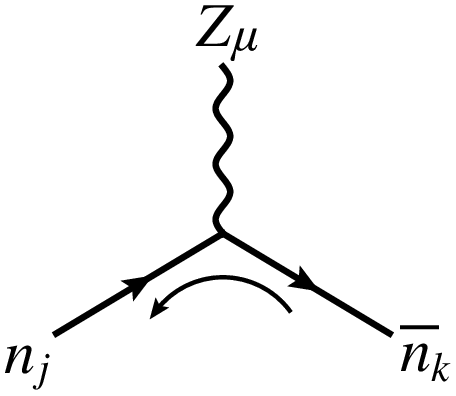}
\end{gathered}
=C\,\Gamma^{\mu{\rm T}}_{jk}C^{-1},
\end{equation}
where $\Gamma^{\mu{\rm T}}_{jk}$ is the transpose of $\Gamma^{\mu}_{jk}$.
To complete the set of one-loop diagrams $ZZZ^*$, BS must be implemented to each of the diagrams of Fig.~\ref{Feyndiags}, which yields a total of 24 contributing generic diagrams. 
\\

The amplitude to calculate is given by
\begin{equation}
ie\,\Gamma^{\nu}_{\alpha\beta\mu}=\sum_{i=1}^6\sum_{j=1}^6\sum_{k=1}^6ie\,\Gamma^{ijk}_{\alpha\beta\mu},
\end{equation}
with the partial-amplitude contribution $ie\,\Gamma^{ijk}_{\alpha\beta\mu}$ diagramatically expressed as
\begin{widetext}
\begin{equation}
ie\,\Gamma^{ijk}_{\alpha\beta\mu}=\hspace{0.2cm}
\begin{gathered}
\vspace{-0.2cm}
\includegraphics[width=2cm]{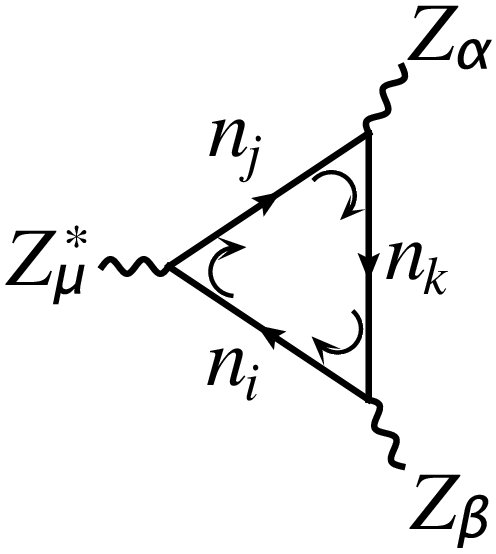}
\end{gathered}
+
\begin{gathered}
\vspace{-0.2cm}
\hspace{0.2cm}
\includegraphics[width=2cm]{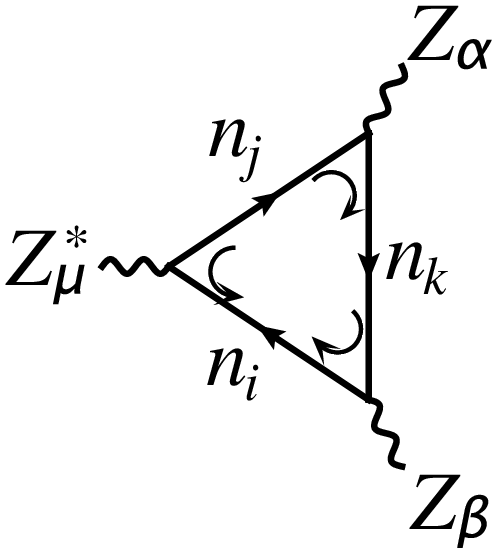}
\end{gathered}
+
\begin{gathered}
\vspace{-0.2cm}
\hspace{0.2cm}
\includegraphics[width=2cm]{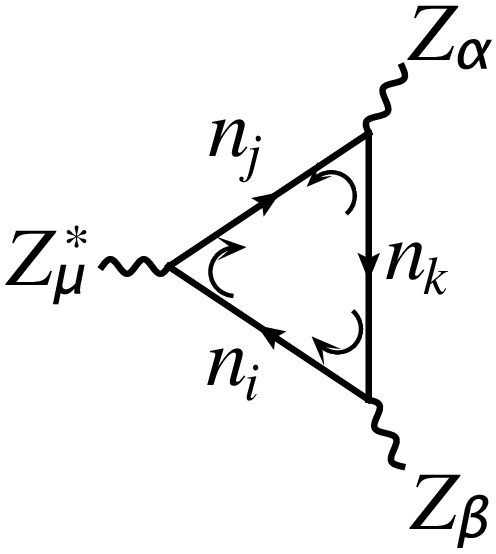}
\end{gathered}
+
\begin{gathered}
\vspace{-0.2cm}
\hspace{0.2cm}
\includegraphics[width=2cm]{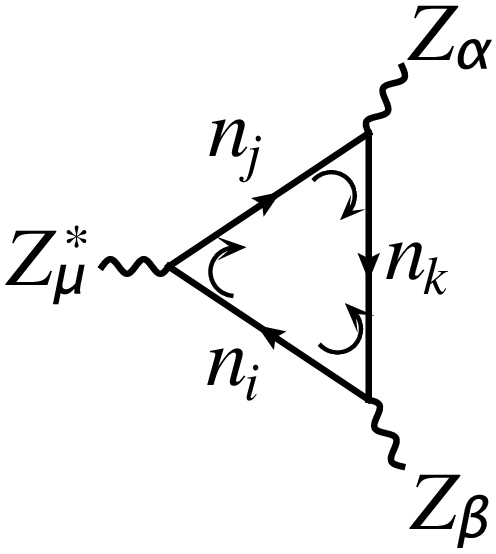}
\end{gathered}
+
\textrm{BS diagrams}.
\label{sumofdiags}
\end{equation}
\end{widetext}
The superficial degree of divergence of any of the contributing diagrams explicitly shown in Eq.~(\ref{sumofdiags}), as well as of those obtained from BS, is 1, so any of them might  bear UV divergences. Keep in mind, however, that, in view of the absence of a coupling $ZZZ$ at the tree level, the total contribution $\Gamma^\nu_{\alpha\beta\mu}$ is expected to be UV finite. In order to give these latent divergences in the amplitude a proper treatment, we use the method of dimensional regularization~\cite{BoGi,tHVe}, so the amplitude is set in $D$ spacetime dimensions with $D$ a complex number such that $D\to4$. In this context, $\int\frac{d^4k}{(2\pi)^4}$ is replaced by $\mu^{4-D}\int\frac{d^Dk}{(2\pi)^D}=\frac{i}{(4\pi)^2}\frac{(2\pi\mu)^{4-D}}{i\pi^2}\int d^Dk$, in loop integrals, where $\mu$ is the renormalization scale. The algebraic procedure to calculate the amplitude is executed by following the tensor-reduction method~\cite{PaVe,DeSt}, which we implement through the software tools FeynCalc~\cite{Feyncalc1,Feyncalc2,Feyncalc3} and Package-X~\cite{PackageX}. After data processing, we get a vertex-function partial contribution with Lorentz-covariant structure
\begin{eqnarray}
&&
\Gamma^{ijk}_{\alpha\beta\mu}=\eta_1^{ijk}\big( p_\alpha\,g_{\beta\mu}+p_\beta\,g_{\alpha\mu} \big)
\nonumber \\ &&
\hspace{0.5cm}
+\eta_2^{ijk}(q^\rho_1-q^\rho_2){\rm tr}\{ \gamma_\mu\gamma_\beta\gamma_\alpha\gamma_\rho\gamma_5 \}
\nonumber \\ &&
\hspace{0.5cm}
+\eta_3^{ijk}(q_1^\sigma-q_2^\sigma) p^\rho\big( {\rm tr}\{ \gamma_\mu\gamma_\alpha\gamma_\rho\gamma_\sigma\gamma_5 \}q_{1\beta}
\nonumber \\ &&
\hspace{0.5cm}
-{\rm tr}\{ \gamma_\mu\gamma_\beta\gamma_\rho\gamma_\sigma\gamma_5 \}q_{2\alpha} \big),
\label{partialAmp}
\end{eqnarray}
where $p_\mu$ terms have been disregarded, in conformity with the discussion of Section~\ref{model}. Here, the factors $\eta_1^{ijk}$, $\eta_2^{ijk}$, and $\eta_3^{ijk}$ are functions on neutrino masses $m_{n_j}$, the $Z$-boson mass $m_Z$, and $s=p^2$. They are given in terms of 1-point, 2-point, and 3-point Passarino-Veltman scalar functions, which are defined as~\cite{PaVe,tHVescalar}
\begin{widetext}
\begin{eqnarray}
A_0\big( m^2_0 \big)&=&\frac{(2\pi\mu)^{4-D}}{i\pi^2}\int d^Dk\frac{1}{k^2-m_0^2},
\label{A0def}
\\ 
B_0\big( p_1^2,m_0^2,m_1^2 \big)&=&\frac{(2\pi\mu)^{4-D}}{i\pi^2}\int d^Dk\frac{1}{\big( k^2-m_0^2 \big)\big( (k+p_1)^2-m_1^2 \big)},
\label{B0def}
\\ 
C_0\big( p_1^2,(p_1-p_2)^2,p_2^2,m_0^2,m_1^2,m_2^2 \big)&=&\frac{(2\pi\mu)^{4-D}}{i\pi^2}\int d^Dk\frac{1}{\big( k^2-m_0^2 \big)\big( (k+p_1)^2-m_1^2 \big)\big( (k+p_2)^2-m_2^2 \big)}.
\label{C0def}
\end{eqnarray}
\end{widetext}
The factors $\eta_X^{ijk}$, with $X=1,2,3$, depend, in particular, on scalar functions $A_0(m_{n_j})$, $B_0(m_Z^2,m_{n_j}^2,m_{n_k}^2)$, $B_0(s,m_{n_j}^2,m_{n_k}^2)$, and $C_0(m_Z^2,m_Z^2,s,m_{n_j}^2,m_{n_k}^2,m_{n_i}^2)$, with all the scalar functions involving all possible combinations of neutrino masses $m_{n_j},m_{n_k},m_{n_i}$ in their arguments. 
\\

Dimensional regularization is the approach most commonly used to tackle UV-divergent loop integrals, as this scheme is suitable for software implementation. Moreover, preservation of gauge invariance is customarily argued to be an appealing feature of this regularization method. Nonetheless, problems may arise when calculations involve the chirality matrix $\gamma_5$, which is incompatible with dimensional regularization, as it has been nicely discussed in Ref.~\cite{Jegerlenhner}. In fact, this incompatibility can generate spurious anomalous contributions, potentially able to spoil Ward identities. Paths to deal with dimensionally-regularized chiral amplitudes have been proposed. In the naive dimensional regularization the $\gamma_5$ is assumed to fulfill $\{ \gamma^\mu,\gamma_5 \}=0$, where $\gamma^\mu$ is any of the $D$ gamma matrices. In contraposition, the \textquoteleft t Hooft-Veltman approach~\cite{tHVe} works under the assumption that the chirality matrix anticommutes with $\gamma^0, \gamma^1, \gamma^2,\gamma^3$, but commutes with the remaining $D-4$ Dirac matrices. Variants of the \textquoteleft t Hooft-Veltman way can also be found~\cite{BrMa,CFH,AoTo,Bonneau}. A main issue of the occurrence of the chirality matrix  in calculations executed in the dimensional regularization approach are traces ${\rm tr}\{ \gamma^\mu\gamma^\nu\gamma^\rho\gamma^\lambda\gamma_5 \}$, which are inconsistently set to 0 when working in the framework of naive dimensional regularization. In \textquoteleft t Hooft-Veltman-like treatments, these traces are nonzero, but illegitimate terms, such as the aforementioned fake anomalies, might emerge. Therefore, calculations of amplitudes which involve this sort of traces must be worked out carefully. With the objective of sensibly dealing with this issue, we have left such traces unevaluated, as it can be seen in Eq.~(\ref{partialAmp}). According to this equation, this sort of traces play a role in the partial amplitude contribution $\Gamma_{\alpha\beta\mu}^{ijk}$, being part of the terms with factors $\eta_2^{ijk}$ and $\eta_3^{ijk}$. 
\\

Any contribution $\eta_X^{ijk}$, in Eq.~(\ref{partialAmp}), can be expressed as
\begin{equation}
\eta_X^{ijk}=\sum_a\tilde{\eta}_aA_0^{(a)}+\sum_b\hat{\eta}_bB_0^{(b)}+\sum_c\bar{\eta}_cC_0^{(c)},
\end{equation}
where $A_0^{(a)}$, $B_0^{(b)}$, and $C_0^{(c)}$ generically denote the different 1-point, 2-point, and 3-point Passarino-Veltman scalar functions featured in the contributions. Each sum in each term of this equation runs over the scalar functions $A^{(a)}_0$, $B^{(b)}_0$, or $C^{(c)}_0$ found in the factor. 
The purpose of this equation is then to sketch the structure of these coefficients, with respect to their Passarino-Veltman scalar-functions dependence. Eq.~(\ref{C0def}) shows that the superficial degree of divergence of 3-point scalar functions is $-2$ for $D=4$, so the $C_0$'s are finite in the UV sense. On the other hand, inspection of the definitions given in Eqs.~(\ref{A0def}) and (\ref{B0def}) leads to the conclusion that 1-point and 2-point Passarino-Veltman scalar functions are UV divergent. In fact, these functions can be written, in general, as $A_0(m^2)=m\big( \Delta^{\rm div.}+\log\mu^2 \big)+A_0^{\rm fin.}$ and $B_0=\Delta^{\rm div.}+\log\mu^2+B_0^{\rm fin.}$, where $m$ is some mass, $\Delta^{\rm div.}$ is a factor which diverges as $D\to4$, and $A_0^{\rm fin.}$, $B_0^{\rm fin.}$ are both finite contributions in the limit as $D\to4$. Note that the divergent factor $\Delta^{\rm div.}$ is shared by all the 1-point and 2-point scalar functions, no matter which their momentum and mass arguments are, so a cancellation of divergences may happen. It turns out that this indeed the case, so all the factors $\eta_X^{ijk}$ are free of UV divergences. Therefore, the limit $D\to4$ can be taken and the remaining traces in Eq.~(\ref{partialAmp}) can be straightforwardly evaluated. For starters, we have ${\rm tr}\{ \gamma_\mu\gamma_\beta\gamma_\alpha\gamma_\rho\gamma_5 \}=-4i\epsilon_{\mu\rho\sigma\lambda}$. Furthermore, with the aid of the Schouten identity~\cite{NeeVe}, we find that $(q_1^\sigma-q_2^\rho)p^\rho\big( {\rm tr}\{ \gamma_\mu\gamma_\alpha\gamma_\rho\gamma_\sigma\gamma_5 \}q_{1\beta}-{\rm tr}\{ \gamma_\mu\gamma_\beta\gamma_\rho\gamma_\sigma\gamma_5 \}q_{2\alpha} \big)=-4i\epsilon_{\mu\alpha\beta\rho}(q_1^\rho-q_2^\rho)s$, which allows us to cast Eq.~(\ref{partialAmp}) into the parametrization given in Eq.~(\ref{ZZZvfunction}), once $p_\mu$ terms are neglected and transversality conditions implemented. Then, the identifications
\begin{eqnarray}
&&
f_4=\frac{-im_Z^2}{s-m_Z^2}\sum_{i=1}^6\sum_{j=1}^6\sum_{k=1}^6\eta^{ijk}_1,
\label{f4summed}
\\ &&
f_5=\frac{-4m_Z^2}{s-m_Z^2}\sum_{i=1}^6\sum_{j=1}^6\sum_{k=1}^6\big( \eta^{ijk}_2-s\,\eta_3^{ijk} \big),
\label{f5summed}
\end{eqnarray}
of the CP-odd form factor $f_4$ and the CP-even form factor $f_5$, is directly made, in accordance with the $ZZZ^*$ parametrization shown in Eq.~(\ref{ZZZvfunction}). 
\\

\section{Estimations and discussion of results}
\label{phenonumeric}
The main objective of the present section is the estimation and analysis of the one-loop contributions from Majorana neutrinos, defined within the framework of Ref.~\cite{Pilaftsis}, to the form factors characterizing the vertex $ZZZ^*$. The one-loop SM contribution to this neutral triple gauge vertex was calculated a couple decades ago in Refs.~\cite{GLR,CDRR}. Such calculations where shown to contribute to the CP-even form factor $f_5$, whereas CP-nonpreserving contributions, associated to $f_4$, were found to be absent. In these works, contributions were analyzed for different values of $\sqrt{s}=\sqrt{p^2}$, showing that the SM yields CP-even effects within ${\cal O}(10^{-4}) - {\cal O}(10^{-3})$. BSM physics has also been considered as a source of $ZZZ^*$ contributions. In fact, the same Refs.~\cite{GLR,CDRR} deal with contributions from the Minimal Supersymmetric SM. Moreover, SM extensions with two Higgs doublets defined the scenarios considered by the authors of Refs.~\cite{CKP,BFFRS} to calculate contributions to CP violation in $ZZZ^*$. As another instance, non-minimal extended scalar sectors, featuring several Higgs multiplets characterized by nondiagonal couplings of the $Z$ boson to charged Higgs fields, were explored by the authors of Ref.~\cite{MTT}, who aimed at the generation of CP-odd contributions to $ZZZ^*$. In Ref.~\cite{DGM}, little-Higgs models were the framework within which the $ZZZ^*$ vertex was calculated at one loop. The model-independent approach provided by the formalism of effective Lagrangians~\cite{LLR,BuWy,Wudka,DGMP} can be used to address BSM physics by assuming the underlying new-physics formulation to govern nature at an energy scale lying far away from the electroweak scale. Investigations of the neutral TGC $ZZZ$, carried out in Refs.~\cite{LPTT,Degrande,Alcaraz}, have profited from such a general formalism.
\\

Processes occurring in electron-positron colliders include $e^+e^-\to ZZ$, in which the neutral TCG $ZZZ^*$ participates though $s$-channel loop diagrams with a virtual $Z$ boson produced by the initial-state electron-positron pair. The Large Electron-Positron Collider, better know as LEP, is the most powerful $e^+e^-$ colliding machine ever been built. Even though the LEP ceased to operate since 2000, a subsequent high-precision analysis of TGCs from data collected at a CME ranging within $130\,{\rm GeV}-209\,{\rm GeV}$, taken by the LEP's four detectors, ALEPH, DELPHI, L3 and OPAL, was carried out in Ref.~\cite{LEP2013TGC}. This study reported agreement with SM expectations, while it reached combined upper limits, of order $10^{-1}$, on both $ZZZ$ form factors $f_4$ and $f_5$. The construction of more powerful $e^+e^-$ colliders, aimed at higher-precision studies, are part of the experimental agenda. Among the anticipated next-generation colliders of this kind, we have the International Linear Collider~\cite{ILCandLHC,ILCtechnicalreport} (ILC), the CERN Compact Linear Collider~\cite{CLIC}, and the Circular Electron-Positron Collider~\cite{CEPC}. While a number of estimations on the sensitivity of next-generation electron-positron colliders to the TGCs $WW\gamma$ and $WWZ$ are available~\cite{ILCandLHC,ILCtechrepTGC,TGCatCEPC,BKGH}, not much has been said regarding the sensitivity of such kind of machines to neutral gauge couplings. In Ref.~\cite{RaSi}, $Z$-boson polarization asymmetries in the processes $e^+e^-\to ZZ$ and $e^+e^-\to Z\gamma$ were considered in order to address sensitivity of some future $e^+e^-$ collider to neutral TCGs. The authors of that paper arrived at the conclusion that a next-generation electron-positron collider working at CME of $500\,{\rm GeV}$ and with an integrated luminosity of $100\,{\rm fb}^{-1}$ would be able to establish upper limits of order $\sim10^{-3}$ on all neutral TGCs. Hadron colliders have been also used to probe neutral TGCs. The D0 experiment, at the Fermilab Tevaron Collider, was able to give bounds as restrictive as $\sim10^{-1}$ on the $ZZZ$ and $ZZ\gamma$ couplings by using data taken from $p\bar{p}$ collisions at a CME of $1.96\,{\rm TeV}$~\cite{TevatronZZZbest}. Nonetheless, the nowadays best limits on the $ZZZ$ coupling have been given by the CMS Collaboration, of the Large Hadron Collider, which established, in Ref.~\cite{CMSbestZZZ}, the bounds
\begin{eqnarray}
-6.6\times10^{-4}<f_4<6.0\times10^{-4},
\\
-5.5\times10^{-4}<f_5<7.5\times10^{-4},
\end{eqnarray}
which were determined from data on $pp$ collisions at a CME of $13\,{\rm TeV}$, with an integrated luminosity of $137\,{\rm fb}^{-1}$. Great relevance is bore by these limits, as they are of the same order as the SM prediction~\cite{GLR,CDRR}. 
\\

Recall Eq.~(\ref{xidef}), which defines the $3\times3$ matrix $\xi$, and then note that this matrix is complex and quite general, only restricted by the conditions $|\xi_{jk}|<1$, fulfilled by all its components. For the sake of practicality, aiming at an estimation of the $ZZZ^*$ contributions whose analytical calculation was discussed throughout Section~\ref{phenoanalytic}, we follow Ref.~\cite{MMNS}, where this matrix was expressed as
\begin{equation}
\xi=\hat{\rho}X.
\label{xieqsrhoX}
\end{equation}
Here, $\hat{\rho}$ is a real and positive number which equals the modulus of the entry $\xi_{jk}$ with the largest magnitude. In this context, the constraint $\hat{\rho}<1$ holds. Furthermore, $X$ is a $3\times3$ complex matrix whose largest entry has modulus 1. The matrix ${\cal C}$, previously given in terms of $\xi$ in Eq.~(\ref{Cinxi}), is thus written as
\begin{equation}
{\cal C}\simeq
\left(
\begin{array}{cccc}
{\bf 1}_3-\hat{\rho}^2XX^\dag &&& \hat{\rho}X\big( {\bf 1}_3-\hat{\rho}^2X^\dag X \big)
\vspace{0.4cm}
\\
\hat{\rho}\big( {\bf 1}_3-\hat{\rho}^2X^\dag X \big)X^\dag &&& \hat{\rho}^2X^\dag X
\end{array}
\right).
\end{equation}
The investigation performed in Ref.~\cite{MMNS}, which featured the authors of the present paper, explored the contributions from Majorana neutrinos to the vertex $WW\gamma$, at one loop. In that work, the values $\hat\rho=0.58$ and $\hat\rho=0.65$ were found to allow for contributions barely within ILC expected sensitivity at a CME of $\sqrt{s}=800\,{\rm GeV}$. Taking that work as a reference, in what follows the value $\hat\rho=0.65$ is used for the estimations and analyses of the present paper. At this point, it should be mentioned that the CMS Collaboration carried out a remarkable model-independent analysis of heavy-neutrino masses in which upper limits on $|{\cal B}_{eN_k}|^2$ and $|{\cal B}_{\mu N_k}|^2$, defined by us in Eqs.~(\ref{Banu}) and (\ref{BaN}), were determined for different values of some heavy-neutrino mass $m_{N_k}$~\cite{CMSneutrinomass}. The results of that paper are displayed in graphs plotted in the parameter spaces $(m_{N_k},|{\cal B}_{eN_k}|^2)$ and $(m_{N_k},|{\cal B}_{\mu N_k}|^2)$. According to the $|{\cal B}_{eN_k}|^2$ graph, the aforementioned value $\hat\rho=0.65$ is consistent with masses $m_{N_k}\gtrsim850\,{\rm GeV}$, whereas the $|{\cal B}_{\mu N_k}|^2$ graph allows for this $\hat\rho$ value if $m_{N_k}\gtrsim1000\,{\rm GeV}$ holds.
\\

We find it worth emphasizing that one-loop contributions to $f_4$ and $f_5$ from virtual neutrinos are always complex valued. To understand this statement, note first that any vertex in any diagram of Fig.~\ref{Feyndiags} connects a $Z$-boson line with a couple of loop neutrino lines. Whenever, for some $j$ and $k$, the condition $m_Z>m_{n_j}+m_{n_k}$, among the masses of the field lines involved in a vertex, hold, the resulting analytic expression for the Feynman diagram turns out to be complex valued. And the same goes for any vertex with a virtual $Z$-boson line as long as $\sqrt{s}>m_{n_j}+m_{n_k}$ is fulfilled. On the contrary, if all the vertices in some contributing diagram are such that $m_Z<m_{n_j}+m_{n_k}$ and $\sqrt{s}<m_{n_j}+m_{n_k}$, whichever $j$ and $k$ are, the resulting analytic expression is real. Then observe that the multiple sums given in Eqs.~(\ref{f4summed}) and (\ref{f5summed}) always come along with diagrams involving vertices $Z\nu_j\nu_k$, coupling a $Z$ boson field with two light neutrinos, in which case $m_Z>m_{\nu_j}+m_{\nu_k}$ happens, thus yielding imaginary-part contributions.
\\

\subsection{CP-odd contributions}
We start our discussion by considering the CP-odd $ZZZ^*$ contributions, quantified by the form factor $f_4$. We split each neutrino sum as $\sum_{j=1}^6=\sum_{\nu_j}+\sum_{N_j}$, where $\nu_j$ runs over light-neutrino fields $\nu_1$, $\nu_2$, $\nu_3$, whereas $N_j$ does it over the three heavy-neutrino fields $N_1$, $N_2$, $N_3$. Then, the triple sum in Eq.~(\ref{f4summed}) is written as
\begin{eqnarray}
&&
\sum_{i=1}^6\sum_{j=1}^6\sum_{k=1}^6=\sum_{\nu_i,\nu_j,\nu_k}+\sum_{\nu_i,\nu_j,N_k}+\sum_{\nu_i,N_j,\nu_k}+\sum_{N_i,\nu_j,\nu_k}
\nonumber \\ && \hspace{0.5cm}
+\sum_{N_i,N_j,\nu_k}+\sum_{N_i,\nu_j,N_k}+\sum_{\nu_i,N_j,N_k}+\sum_{N_i,N_j,N_k}.
\label{sumCPoddcontributions}
\end{eqnarray}
We have verified that taking, in a contributing diagram, the three virtual-neutrino masses the same, that is $m_{n_i}=m_{n_j}=m_{n_k}$, renders the corresponding contribution to $f_4$ zero. This means that CP-odd contributions from diagrams involving only light neutrinos are expected to be quite suppressed, and the same goes for diagrams in which only heavy neutrinos participate, for the neutrino model under consideration requires the spectrum of heavy-neutrino masses to be quasi-degenerate~\cite{Pilaftsis}. In this context, any CP-odd significant contribution is expected to emerge from diagrams in which both light and heavy neutrinos participate, so terms in $f_4$ with triple sums $\sum_{\nu_i}\sum_{\nu_j}\sum_{\nu_k}$ and $\sum_{N_i}\sum_{N_j}\sum_{N_k}$ are from here on disregarded. 
\\

We have found that the occurrence of the CP-odd contribution, $f_4$, requires the matrix $\xi$ to be complex, while such an effect vanishes if this matrix is real or imaginary. Therefore, as inferred from Eq.~(\ref{xieqsrhoX}), the matrix $X$ must be complex, with $\mathfrak{Re}( X )\ne0$ and $\mathfrak{Im}( X )\ne0$. Taking a pragmatic approach, we use $X=e^{i\phi}\cdot{\bf 1}_3$. Even though this form of $X$ is by no means a general texture, it allows us to get an estimation of the CP-violating contributions while avoiding a large number of unknown parameters. Nonetheless, let us emphatically point out that we have tried matrix textures other than ${\bf 1}_{3}$, but found no significant variations in our numerical estimations. Now we take the approximation that $m_{\nu_1}\approx m_\ell$, $m_{\nu_2}\approx m_\ell$, $m_{\nu_3}\approx m_\ell$, with $\ell$ labeling ``light''. Furthermore, in accordance with Ref.~\cite{Pilaftsis}, the heavy-neutrino mass spectrum is restricted to be quasi-degenerate, so $m_{N_1}\approx m_h$, $m_{N_2}\approx m_h$, and $m_{N_3}\approx m_h$ are assumed, where $h$ stands for ``heavy''. In this context, the CP-odd form-factor contribution $f_4$ is expressed as
\begin{widetext}
\begin{eqnarray}
&&
f_4\approx\frac{9\alpha\,m_\ell m_h\,\hat{\rho}^2 \big(\hat{\rho} ^2-1\big)^2\sin 2\phi }{\pi s \big( s-m_Z^2 \big)\big( s-4m_Z^2 \big)\sin^3 2\theta_{\rm W}}
\bigg[2m_Z^2\sqrt{m_\ell^4-2\big(m_h^2+s\big)m_\ell^2+\big(m_h^2-s\big)^2} \big(2 \hat{\rho}^2-1\big) \log\left\{ \frac{g(s,m^2_\ell,m^2_h)}{m_\ell m_h} \right\}
\nonumber \\ &&
\hspace{2cm}
-2m_Z^2 \sqrt{s \big(s-4 m_\ell^2\big)}\bigg( \big(\hat{\rho} ^2-1\big) \log\left\{\frac{g(s,m^2_\ell,m^2_\ell)}{m^2_\ell}\right\}
+ \hat{\rho} ^2 \log\left\{\frac{g(s,m^2_h,m^2_h)}{m_h^2}\right\} \bigg)
\nonumber \\ &&
\hspace{2cm}
-2\sqrt{m_\ell^4-2\big(m_h^2+m_Z^2\big)m_\ell^2+\big(m_h^2-m_Z^2\big)^2} \left(2m_Z^2-s\right) \big(2 \hat{\rho}^2-1 \big) \log\left\{ \frac{g(m^2_Z,m^2_\ell,m^2_h)}{m_\ell m_h} \right\}
\nonumber \\ &&
\hspace{2cm}
+2m_Z\sqrt{m_Z^2-4 m_\ell^2 } \big(2m_Z^2-s\big)\bigg( \big(\hat{\rho}^2-1\big) \log\left\{ \frac{g(m^2_Z,m^2_\ell,m^2_\ell)}{m_\ell^2} \right\}+\hat{\rho}^2\log\left\{ \frac{g(m^2_Z,m^2_h,m^2_h)}{m_h^2} \right\} \bigg)
\nonumber \\ &&
\hspace{2cm}
-\frac{\big(4 m_Z^4-5m_Z^2\,s+s^2\big)\big(\big(2 \hat{\rho} ^2-1\big) \big( m_\ell^2-m_h^2 \big)+2m_Z^2 \big)}{s-4 m_Z^2} \log\left\{ \frac{m_\ell^2}{m_h^2} \right\}
\nonumber \\ &&
\hspace{2cm}
+m_Z^2\,s\big(2 m_\ell^2-2 m_h^2+2m_Z^2-s\big)\big(\hat{\rho}^2-1\big)C^{(\ell,h,\ell)}_0
-m_Z^2\left(2 m_Z^2-s\right)\left(2 m_\ell^2-2 m_h^2+s\right)\big(\hat{\rho}^2-1\big)C^{(h,\ell,\ell)}_0
\nonumber \\ &&
\hspace{2cm}
+m_Z^2\,s\big( -2 m_\ell^2+2 m_h^2+2m_Z^2-s \big) \hat{\rho}^2C^{(h,\ell,h)}_0
-m_Z^2\big(2 m_Z^2-s\big)\big(-2 m_\ell^2+2 m_h^2+s\big)\hat{\rho}^2C^{(h,h,\ell)}_0
\bigg],
\label{f4explicit}
\end{eqnarray}
\end{widetext}
where
\begin{eqnarray}
&&
g(m^2,m^2_1,m^2_2)=\frac{1}{2}\Big( m_1^2+m_2^2-m^2
\nonumber \\ &&
\hspace{0.5cm}
+\sqrt{m_1^4-2 m_1^2\left(m^2+m_2^2\right)+\left(m_2^2-m^2\right)^2}\Big)
\end{eqnarray}
has been defined. Moreover, $\alpha$ is the fine structure constant. We have also used the notation
\begin{eqnarray}
C_0^{(n_1,n_2,n_3)}=C_0(m_Z^2,m_Z^2,s,m^2_{n_1},m^2_{n_2},m^2_{n_3}),
\end{eqnarray}
for 3-point scalar functions, with the sole purpose of getting a more compact expression. To write down Eq.~(\ref{f4explicit}), the 1-point and 2-point scalar functions, the $A_0$'s and the $B_0$'s, have been solved explicitly. In the process, all UV divergences have been cancelled and the limit as $D\to4$ has been taken. While 3-point functions $C_0$ remain indicated in this equation, keep in mind that they are UV finite. Note that the whole expression for $f_4$ is proportional to $\sin2\phi$, with $\phi$ the phase earlier introduced in the considered texture for the matrix $X$. Then, Eq.~(\ref{f4explicit}) illustrates how rendering the $X$ matrix real or imaginary, by taking $\phi=0,\frac{\pi}{2},\pi,\frac{3\pi}{4}$, yields the complete elimination of $f_4$. On the other hand, optimal values for this phase, in the sense that they do not introduce any suppression to the contribution $f_4$, are $\phi=\frac{\pi}{4},\frac{3\pi}{4},\frac{5\pi}{4},\frac{7\pi}{4}$, since in such cases $\sin2\phi=\pm1$.
\\

\begin{figure}[ht]
\centering
\includegraphics[width=8.6cm]{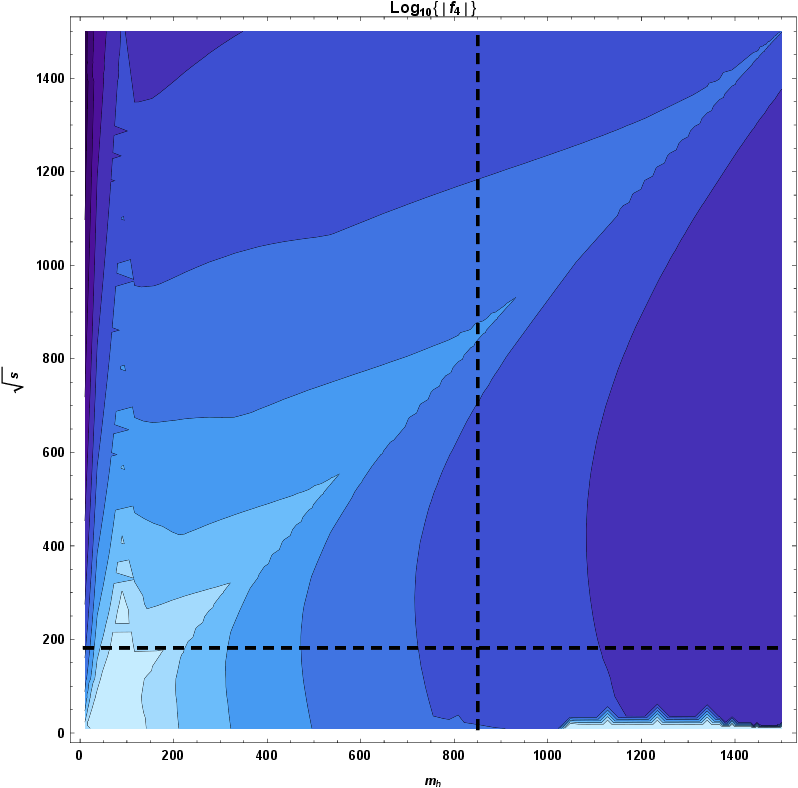}
\includegraphics[width=8.4cm]{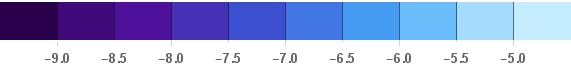}
\caption{\label{mNvsQ2Dgraph} Majorana-neutrino contributions to $\log_{10}|f_4|$ in the region defined by $10\,{\rm GeV}\leqslant m_h\leqslant1500\,{\rm GeV}$ and $10\,{\rm GeV}\leqslant\sqrt{s}\leqslant1500\,{\rm GeV}$, within the parameter space $(m_h,\sqrt{s})$. The horizontal dashed line, at $\sqrt{s}=2m_Z$, represents the threshold for production of $Z$ pairs by $e^+e^-\to ZZ$. The dashed vertical line indicates the heavy-neutrino mass value $m_h=850\,{\rm GeV}$, beyond which our choice $\hat\rho=0.65$ is consistent, in accordance with Ref.~\cite{CMSneutrinomass}.}
\end{figure}

As we just discussed, a few paragraphs ago, $f_4$ is a complex quantity. In this context, we consider, for our forthcoming discussion, the modulus $|f_4|$. We refer the reader to the graph in Fig.~\ref{mNvsQ2Dgraph},
which displays disjoint regions corresponding to different values of $|f_4|$, plotted in the $\big( m_h,\sqrt{s} \big)$ parameter space. Considered values of the heavy-neutrino mass $m_h$ and the CME $\sqrt{s}$ range within $10\,{\rm GeV}\leqslant m_h\leqslant1500\,{\rm GeV}$ and $10\,{\rm GeV}\leqslant\sqrt{s}\leqslant1500\,{\rm GeV}$. The CP-violation phase $\phi=\frac{\pi}{4}$ has been taken because this pick yields optimal contributions, so the values reported here should be rather understood as upper bounds in the sense that different choices for the CP phase $\phi$ would introduce a suppression on the $f_4$ contribution. The $|f_4|$ values plotted in Fig.~\ref{mNvsQ2Dgraph} are given in base 10 logarithmic scale, so that this graph allows one to better appreciate the orders of magnitude of the contributions corresponding to each one of the regions shown. The color scheme of the graph has been set in such a way that the lighter the tone of the region, the larger the $|f_4|$ contribution, which is indicated by the labeling bar below the graph. The lightest tone comprehends contributions of order $\gtrsim10^{-5}$. With this mind, notice that the largest contributions to $|f_4|$ gather within the region defined by $10\,{\rm GeV}\lesssim m_N\lesssim250\,{\rm GeV}$ and $10\,{\rm GeV}\lesssim \sqrt{s}\lesssim400\,{\rm GeV}$, though be aware that most of this region corresponds to a CME $\sqrt{s}$ below the $Z$-pair production threshold, which has been indicated in the graph of Fig.~\ref{mNvsQ2Dgraph} by a horizontal dashed line at $\sqrt{s}=2m_Z$. A vertical dashed line, at $m_h=850\,{\rm GeV}$, has also been added to the graph to specify which values of the heavy-neutrino mass are in conformity with the value $\hat\rho=0.65$, considered for our estimations. With this in mind, notice that the relevant region within the graph of Fig.~\ref{mNvsQ2Dgraph} is the upper-right one, beyond these dashed lines.
\\

\begin{figure}[ht]
\centering
\includegraphics[width=8.6cm]{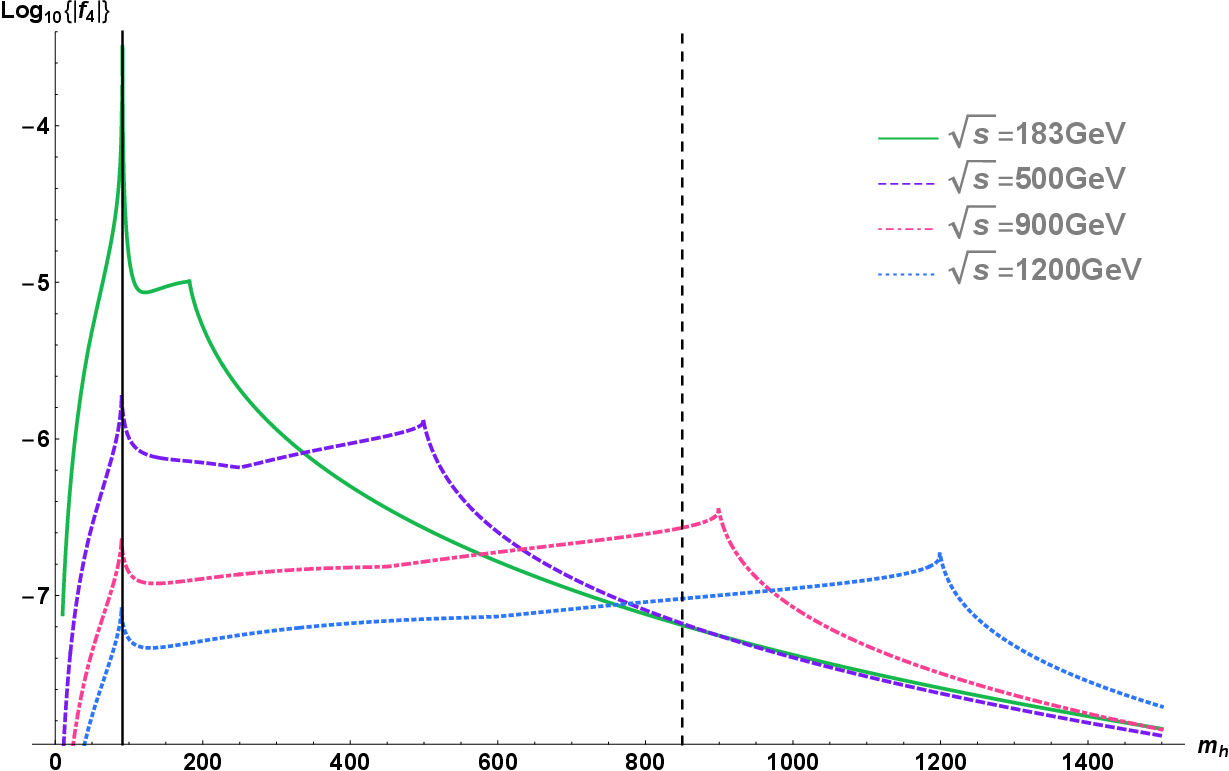}
\vspace{0.5cm}
\\
\includegraphics[width=8.6cm]{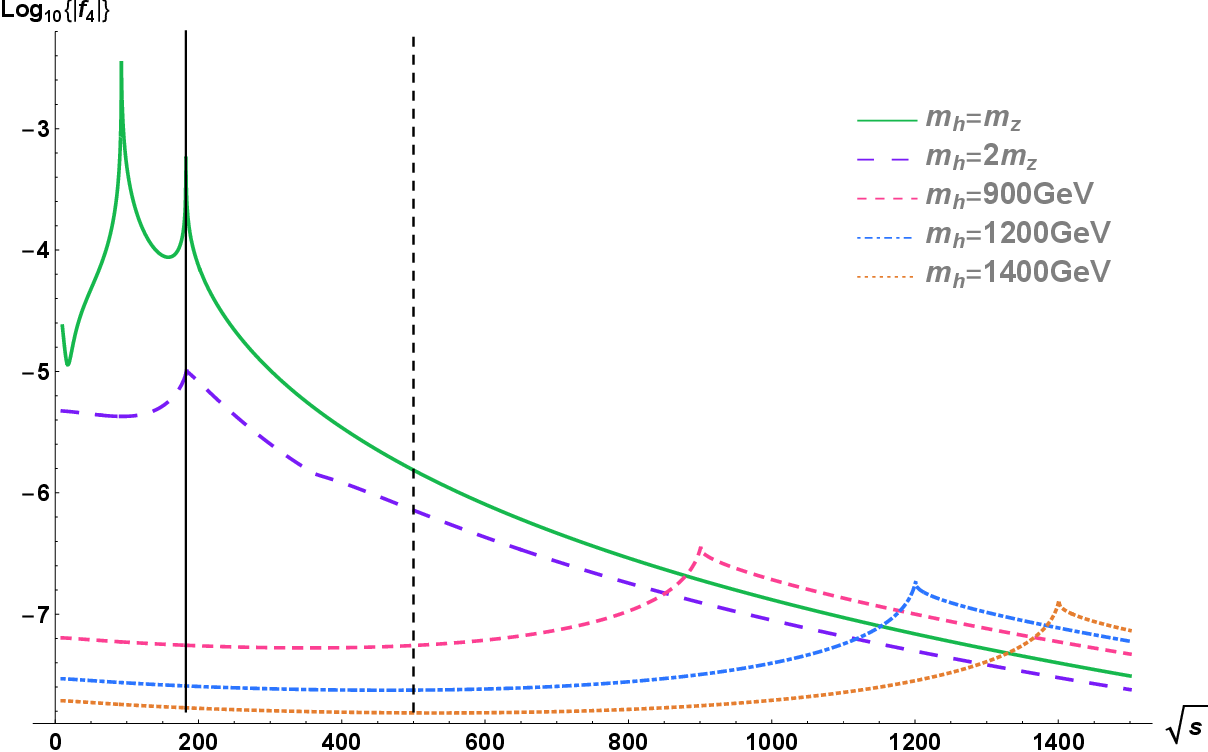}
\caption{\label{2Df4graphs} Upper graph: contributions from Majorana neutrinos to $\log_{10}|f_4|$, as a function on the heavy-neutrino mass $m_h$, for fixed CME values, with the vertical dashed line representing the heavy-neutrino mass value $m_h=850\,{\rm GeV}$. Lower graph: contributions from Majorana neutrinos to $\log_{10}|f_4|$, as a function on the CME $\sqrt{s}$, for fixed values of the heavy-neutrino mass, with the vertical solid line representing the $Z$-pair production threshold and the vertical dashed line indicating the value $\sqrt{s}=500\,{\rm GeV}$.}
\end{figure}

A complementary viewpoint is provided by the graphs of Fig.~\ref{2Df4graphs}.
The upper graph of this figure shows plots of $|f_4|$, in base 10 logarithmic scale, with respect to the heavy-neutrino mass $m_h$, for a variety of fixed values of $\sqrt{s}$. In this graph, $m_h$ ranges from $10\,{\rm GeV}$ to $1500\,{\rm GeV}$, whereas for the CME $\sqrt{s}$ the following values were considered: $\sqrt{s}=183\,{\rm GeV}$, which corresponds to the solid curve; $\sqrt{s}=500\,{\rm GeV}$, represented by the dashed plot; $\sqrt{s}=900\,{\rm GeV}$, used to get the dot-dashed curve; and $\sqrt{s}=1200\,{\rm GeV}$, for the dotted curve. Furthermore, a vertical solid straight line has been added to represent the value $m_h=m_Z$, at which $|f_4|$ has a maximum no matter what the value of $\sqrt{s}$ is. Besides this maximum-valued $|f_4|$ contribution, each curve displays another local maximum, which varies depending on $\sqrt{s}$. Note, however, that such maxima of $|f_4|$ do not necessarily correspond to the relevant largest contributions. For instance, the CME $\sqrt{s}=183\,{\rm GeV}$, just next to the $Z$-pair production threshold, was explored for illustrative purposes and because, according to Fig.~\ref{mNvsQ2Dgraph}, values close to $\sqrt{s}=2m_Z$ yield the largest contributions to $|f_4|$, for certain heavy-neutrino mass values. In fact, for $\sqrt{s}=183\,{\rm GeV}$ a contribution of order $10^{-4}$, of the same order of magnitude as current LHC limits on $f_4$~\cite{CMSbestZZZ}, is produced at $m_N=m_Z$. However, keep in mind that the results reported in Ref.~\cite{CMSneutrinomass} do not allow for a heavy-neutrino mass so small, due to our choice of the $\hat\rho$ parameter. The second maximum for this curve also corresponds to a heavy-neutrino mass value $m_h<850\,{\rm GeV}$. Within the allowed-$m_h$ region, on the other hand, this curve reaches its maximum contribution precisely at $m_h=850\,{\rm GeV}$, which is ${\cal O}(10^{-8})$. The set of relevant maxima for the curves in the upper graph of Fig.~\ref{2Df4graphs}, for $m_h\geqslant850\,{\rm GeV}$, is shown in Table~\ref{f4tablefixedQ},
\begin{table}[ht]
\centering
\begin{tabular}{c|c|c}
\hline\hline
$\sqrt{s}$ & $m_h$ & $|f_4|_{\rm max}$ 
\\ \hline
$183\,{\rm GeV}$ & $850\,{\rm GeV}$ & $6.46\times10^{-8}$
\\ \hline
$500\,{\rm GeV}$ & $850\,{\rm GeV}$ & $6.62\times10^{-8}$
\\ \hline
$900\,{\rm GeV}$ & $850\,{\rm GeV}$ & $2.71\times10^{-7}$
\\ \hline
$900\,{\rm GeV}$ & $898.92\,{\rm GeV}$ & $3.60\times10^{-7}$
\\ \hline
$1200\,{\rm GeV}$ & $850\,{\rm GeV}$ & $9.54\times10^{-8}$
\\ \hline
$1200\,{\rm GeV}$  & $1198.94\,{\rm GeV}$ & $1.86\times10^{-7}$
\\ \hline \hline
\end{tabular}
\caption{\label{f4tablefixedQ} Maximum values of the $|f_4|$ contribution for the curves shown in the upper graph of Fig.~\ref{2Df4graphs}, with the heavy neutrino mass constrained as $m_h\geqslant850\,{\rm GeV}$, as dictated by Ref.~\cite{CMSneutrinomass}.}
\end{table}
where the possibility of having CP-odd contributions as large as ${\cal O}(10^{-7})$ can be appreciated. Such largest contributions correspond to the largest CMEs considered for the graph. At $\sqrt{s}=500\,{\rm GeV}$ (dashed curve), the relevant maxima, corresponding to $m_h=850\,{\rm GeV}$, is ${\cal O}(10^{-8})$, which lies about 5 orders of magnitude below the expected experimental sensitivity estimated in Ref.~\cite{RaSi} for ILC at the same CME
\\

The lower graph of Fig.~\ref{2Df4graphs} displays the $|f_4|$ contribution for five selected fixed values of heavy-neutrino mass $m_h$, as a function on the CME $\sqrt{s}$, which has been varied within $10\,{\rm GeV}\leqslant\sqrt{s}\leqslant1500\,{\rm GeV}$. Again, this graph is given in base 10 logarithmic scale. Regarding our choice of heavy-neutrino masses, we used: $m_h=m_Z$ to get the solid plot; the long-dashes curve was carried out by usage of $m_h=2m_Z$; the heavy-neutrino mass value $m_h=900\,{\rm GeV}$ yielded the short-dashes plot; for the dot-dashed curve, $m_h=1200\,{\rm GeV}$ has been utilized; and $m_h=1400\,{\rm GeV}$ corresponds to the dotted plot. The curves corresponding to $m_h=m_Z$ and $m_h=2m_Z$ have been included for the sole purpose of illustration, as such heavy-neutrino masses are not consistent with $\hat\rho=0.65$, used for our estimations. The graph also includes two vertical lines, of which the solid one refers to the threshold for $Z$-pair production, at $\sqrt{s}=2m_Z$. The vertical dashed line, on the other hand, represents the CME $\sqrt{s}=500\,{\rm GeV}$, used in Ref.~\cite{RaSi} to estimate sensitivity of ILC to neutral TGCs. The maxima associated to the $|f_4|$ contributions for the aforementioned $m_h$ choices, as well as the heavy-neutrino masses yielding such maxima, are displayed in Table~\ref{f4tablefixedmN}.
\begin{table}[ht]
\centering
\begin{tabular}{c|c|c}
\hline \hline
$m_h$ & $\sqrt{s}$ & $|f_4|_{\rm max}$
\\ \hline
$m_Z$ & $2m_Z$ & $7.68\times10^{-4}$
\\ \hline
$2m_Z$ & $183.48\,{\rm GeV}$ & $1.04\times10^{-5}$
\\ \hline
$900\,{\rm GeV}$ & $901.08\,{\rm GeV}$ & $3.59\times10^{-7}$
\\ \hline
$1200\,{\rm GeV}$ & $1201.06\,{\rm GeV}$ & $1.86\times10^{-7}$
\\ \hline
$1400\,{\rm GeV}$ & $1401.05\,{\rm GeV}$ & $1.30\times10^{-7}$
\\ \hline \hline
\end{tabular}
\caption{\label{f4tablefixedmN} Maximum values of the $|f_4|$ contribution for the curves shown in the lower graph of Fig.~\ref{2Df4graphs}.}
\end{table}
At $\sqrt{s}=500\,{\rm GeV}$, the curve given by the choice $m_h=900\,{\rm GeV}$ dominates, though notice that larger masses play the main role at higher CMSs, where contributions of order $10^{-7}$ are generated.
\\

\subsection{CP-even contributions}
By contrast with the contribution $f_4$, discussed in the previous subsection, the CP-preserving contribution $f_5$ does not require the matrix $X$ to be complex in order be nonzero. As we did before, for $f_4$, we take the approximations $m_{\nu_k}\approx m_\ell$ and $m_{N_k}\approx m_h$, for all $k=1,2,3$, in which case we are led to the expression
\begin{widetext}
\begin{eqnarray}
&&
f_5\approx\frac{9\alpha}{2\pi s\big( s-m_Z^2 \big)\big( s-4m_Z^2 \big)^2 \sin^32\theta_{\rm W}}
\bigg[
\xi_1 \log\bigg\{\frac{g\big( m_Z^2,m_h^2,m_h^2 \big)}{ m_h^2}\bigg\}
+\xi_2 \log \bigg\{\frac{g\big( s,m_h^2,m_h^2 \big)}{m_h^2}\bigg\}
+\xi_3\log\bigg\{\frac{m_\ell^2}{m_h^2}\bigg\}
\nonumber \\ &&
+\xi_4 \log\bigg\{\frac{g\big( m_Z^2,m_\ell^2,m_h^2 \big)}{ m_\ell m_h}\bigg\}
+\xi_5\log\bigg\{\frac{g\big( s,m_\ell^2,m_h^2 \big)}{m_\ell m_h}\bigg\}
+\xi_6\log\bigg\{\frac{g\big( m_Z^2,m_\ell^2,m_\ell^2 \big)}{m_\ell^2}\bigg\}
+\xi_7\log\bigg\{\frac{g\big( s,m_\ell^2,m_\ell^2 \big)}{m_\ell^2}\bigg\}
\nonumber \\ &&
+\xi_8\,C_0^{(\ell,\ell,\ell)}
+\xi_9\,C_0^{(h,h,h)}
+\xi_{10}\,C_0^{(h,\ell,h)}
+\xi_{11}\,C_0^{(h,h,\ell)}
+\xi_{12}\,C_0^{(\ell,h,\ell)}
+\xi_{13}\,C_0^{(h,\ell,\ell)}
+\xi_{14}
\bigg].
\label{f5explicit}
\end{eqnarray}
\end{widetext}
This equation displays the expression of $f_5$, once the 1-point and the 2-point Passarino-Veltman scalar functions have been solved, all UV divergencies have been eliminated, and the limit as $D\to4$ has been taken. The expression of the $f_5$ contribution is lengthy, so we present it in a concise manner, in terms of coefficients $\xi_n$, which appear in each term of the equation. The explicit definitions of the coefficients $\xi_n$ can be found in the Appendix. These quantities depend on the mass of the $Z$ boson and on the neutrino masses, $m_\ell$ and $m_h$, as well. The squared CME $s$ is also a variable determining the $\xi_n$ coefficients. Finally, note that the $\xi_n$'s are also functions on the parameter $\hat\rho$ and on the phase $\phi$. Note that the $\phi$-phase dependence does not factorize in $f_5$, as opposite to the CP-odd contribution $f_4$, Eq.~(\ref{f4explicit}). Moreover, notice that usage of the value $\phi=0$, which renders $X$ real, does not eliminate the CP-even contribution, that is, $f_5\big|_{\phi=0}\ne0$.
\\

\begin{figure}[ht]
\center
\includegraphics[width=8.6cm]{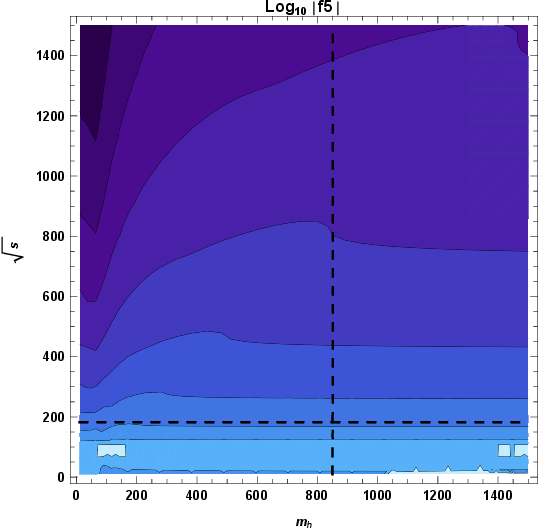}
\includegraphics[width=8.4cm]{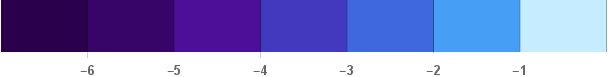}
\caption{\label{mNvsQ2Dgraph2} Majorana-neutrino contributions to $\log_{10}|f_5|$ in the region defined by $10\,{\rm GeV}\leqslant m_h\leqslant1200\,{\rm GeV}$ and $10\,{\rm GeV}\leqslant\sqrt{s}\leqslant1500\,{\rm GeV}$, within the parameter space $(m_h,\sqrt{s})$. The horizontal dashed line, at $\sqrt{s}=2m_Z$, represents the threshold for production of $Z$ pairs by $e^+e^-\to ZZ$. The dashed vertical line indicates the heavy-neutrino mass value $m_h=850\,{\rm GeV}$, beyond which our choice $\hat\rho=0.65$ is consistent, in accordance with Ref.~\cite{CMSneutrinomass}.}
\end{figure}

For our upcoming discussion, we fix the complex phase by $\phi=\frac{\pi}{4}$, which previously yielded an optimal CP-nonconserving contribution $f_4$. This choice has the effect of eliminating a few $f_5$ terms, as they involve the factor $\cos2\phi$. A panorama of the resulting CP-odd contribution $f_5$, in the $(m_h,\sqrt{s})$ parameter space, is given by the graph in Fig.~\ref{mNvsQ2Dgraph2},
which has been effectuated within the region defined by $10\,{\rm GeV}\leqslant m_h\leqslant1500\,{\rm GeV}$ and $10\,{\rm GeV}\leqslant\sqrt{s}\leqslant1500\,{\rm GeV}$. Again, the norm $|f_5|$ has been used. Furthermore, the graph has been plotted in base 10 logarithmic scale, so the different regions comprising it are colored in accordance with the sizes of $\log_{10}\big|f_5\big|$, which depict orders of magnitude of the contributions corresponding to the different points of the parameter space. A labeling bar, beneath the graph, has been added to Fig.~\ref{mNvsQ2Dgraph2} for reference. It shows that lighter tones correspond to larger $|f_5|$, with the largest contributions lying around the $Z$-boson pole $\sqrt{s}=m_Z$. Nevertheless, such sizable contributions are within a region in which the CME is below the threshold for $Z$-boson pair production. Such a threshold has been represented in the graph by a dashed horizontal line, at $\sqrt{s}=2m_Z$. This region has to be disregarded from our discussion, as it plays no role in the physical process under consideration. The vertical straight dashed line, at $m_h=850\,{\rm GeV}$, shows the smallest value of the heavy-neutrino mass $m_h$ which is compatible with $\hat\rho=0.65$, as established by Ref.~\cite{CMSneutrinomass}. Then notice that the region corresponding to $m_h\leqslant850\,{\rm GeV}$ has to be overlooked as well. The resulting relevant region in the $(m_h,\sqrt{s})$ plane turns out to be the one beyond both the $Z$-pair production threshold and $m_h\geqslant850\,{\rm GeV}$. 
\\

\begin{figure}[ht]
\center
\includegraphics[width=8.6cm]{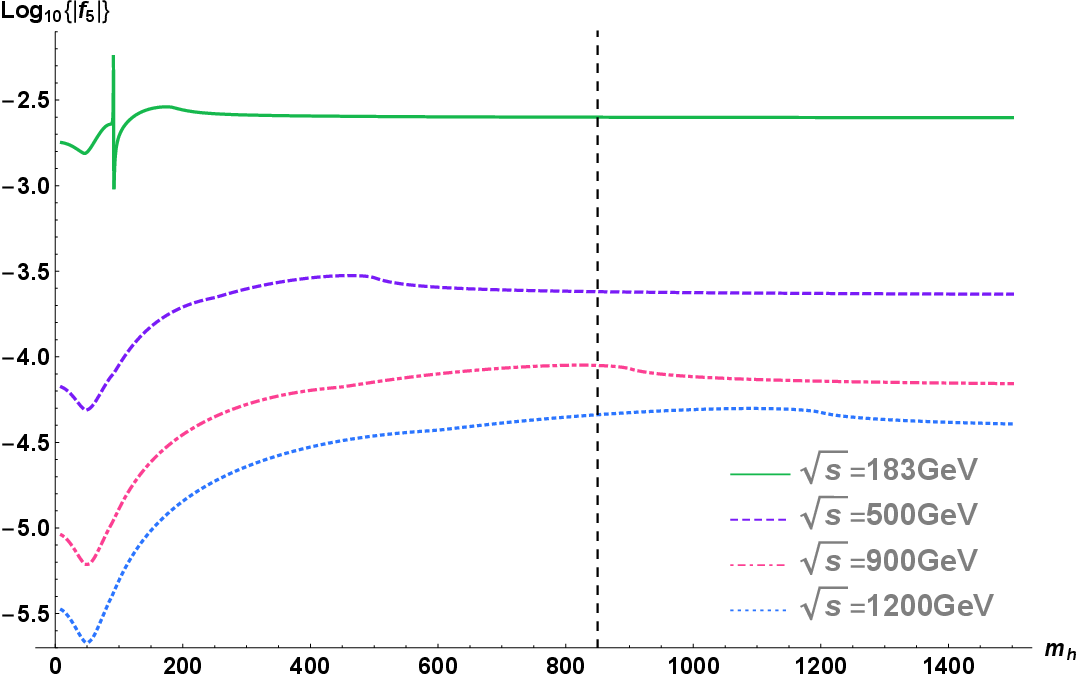}
\vspace{0.5cm}
\\
\includegraphics[width=8.6cm]{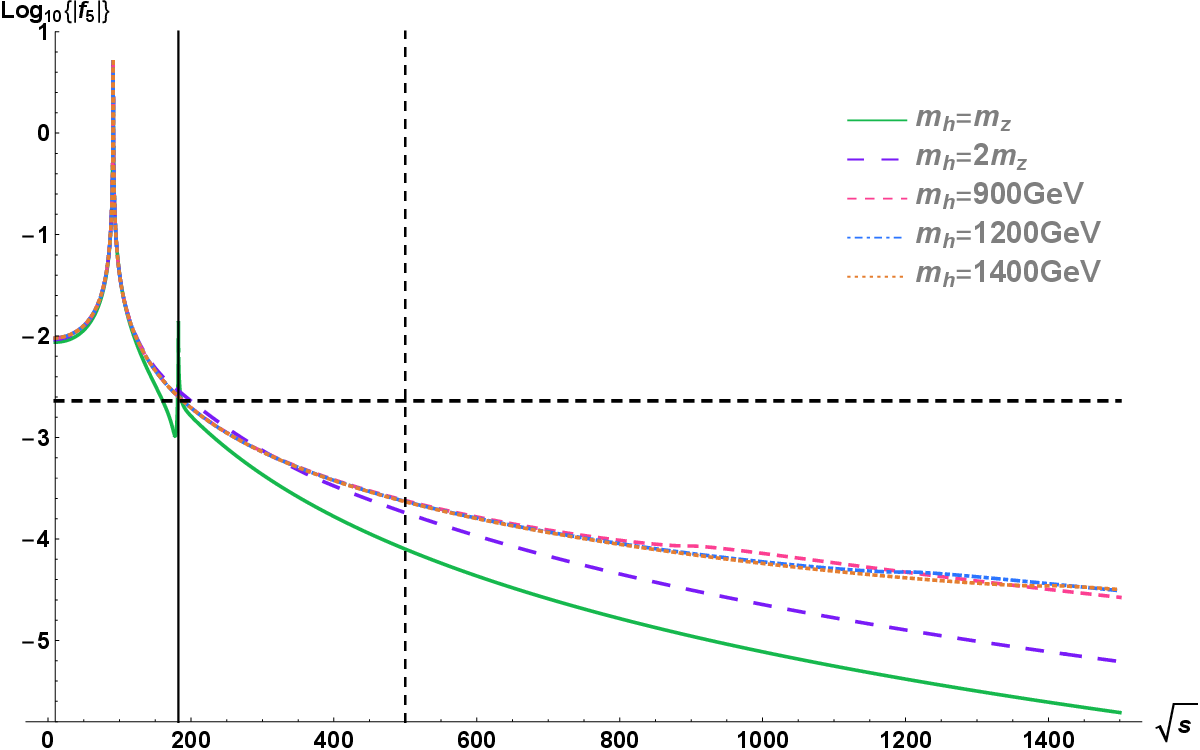}
\caption{\label{2Df5graphs} Upper graph: contributions from Majorana neutrinos to $\log_{10}|f_5|$, as a function on the heavy-neutrino mass $m_h$, for fixed CME values, with the vertical dashed line representing the heavy-neutrino mass value $m_h=850\,{\rm GeV}$. Lower graph: contributions from Majorana neutrinos to $\log_{10}|f_5|$, as a function on the CME $\sqrt{s}$, for fixed values of the heavy-neutrino mass, with the vertical solid line representing the $Z$-pair production threshold and the vertical dashed line indicating the value $\sqrt{s}=500\,{\rm GeV}$.}
\end{figure}

The Majorana-neutrinos contribution to $f_5$ is illustrated by the graphs displayed in Fig.~\ref{2Df5graphs},
in which either $m_h$ or $\sqrt{s}$ is fixed at selected values. The upper graph of this figure shows the behavior of the modulus $|f_5|$, for fixed CME $\sqrt{s}$ values, as a function on the heavy-neutrino mass $m_h$. Here, $\sqrt{s}=183\,{\rm GeV}$, just next to the $Z$-pair production threshold, is represented by the solid plot, whereas the dashed curve corresponds to a CME $\sqrt{s}=500\,{\rm GeV}$. The dot-dashed and the dotted curves stand for $\sqrt{s}=900\,{\rm GeV}$ and $\sqrt{s}=1200\,{\rm GeV}$, respectively. Keep in mind that all the plots have been carried out in base 10 logarithmic scale. A pattern, which can be observed in the region graph of Fig.~\ref{mNvsQ2Dgraph2} but which is more clearly appreciated in the upper graph of Fig.~\ref{2Df5graphs}, is an attenuation of contributions as larger CMEs $\sqrt{s}$ are considered. The investigation carried out in Ref.~\cite{RaSi} yielded an estimation of constraints, from the ILC, on $f_5$ , at a CME of $500\,\rm GeV$. The authors of that work established the restriction $-2.3\times10^{-3}\leqslant f^{\rm ILC}_5\leqslant8.8\times10^{-3}$, based upon which we take $|f^{\rm ILC}_5|\leqslant2.3\times10^{-3}$ for reference. Regarding our calculation, at a CME of $\sqrt{s}=500\,{\rm GeV}$ the contribution (dashed plot, upper graph, Fig.~\ref{2Df5graphs}) varies from $|f_5|_{\rm  min}=4.90\times10^{-5}$, at $m_h=48.53\,{\rm GeV}$, to $|f_5|_{\rm max}=2.99\times10^{-4}$, at $m_h=461.58\,{\rm GeV}$. Nonetheless, this curve reaches its relevant maximum value at $m_h=850\,{\rm GeV}$, with the corresponding contribution amounting to $|f_5|=2.40\times10^{-4}$. Thus, the largest contribution at this choice for the CME lies about one order of magnitude below projected ILC sensitivity. Moreover, according to Ref.~\cite{CDRR}, the SM contribution is $|f_5^{\rm SM}|\approx2.34\times10^{-3}$ at $\sqrt{s}=500\,{\rm GeV}$, so our contribution would be also one order of magnitude below that from the SM. Also recall that the CMS Collaboration has established an upper bound of order $10^{-4}$ on $f_5$~\cite{CMSbestZZZ}.
\\

Another perspective is furnished by the lower graph of Fig.~\ref{2Df5graphs}, where five curves, representing the CP-conserving contribution $|f_5|$, have been plotted in base 10 logarithmic scale, with each one of them determined by a fixed value of the heavy-neutrino mass $m_h$. To this aim, we have chosen the following masses: the solid plot emerges from $m_h=m_Z$; the value $m_h=2m_Z$ has been utilized to generate the long-dashes curve; the heavy-neutrino mass $m_h=900\,{\rm GeV}$ yielded the short-dashes curve; the dot-dashed plot follows from the choice $m_h=1200\,{\rm GeV}$; and, finally, the heavy-neutrino mass $m_h=1400\,{\rm GeV}$ corresponds to the dotted curve. Besides these $m_h$-fixed plots, a horizontal dashed line has been added to the graph to indicate the estimation given by Ref.~\cite{RaSi} of ILC sensitivity to $f_5$. Thus, such a line is given at $|f^{\rm ILC}_5|=2.3\times10^{-3}$. We have also included two vertical lines in this graph, one solid and the other dashed. The solid vertical line represents the threshold for $Z$-pair production, at $\sqrt{s}=2m_Z$. Meanwhile, the dashed vertical line represents the CME value $\sqrt{s}=500\,{\rm GeV}$. Table~\ref{f5fixedmh}
\begin{table}[ht]
\center
\begin{tabular}{c|c|c}
\hline \hline
$m_h$ & $\sqrt{s}$ & $|f_5|$
\\ \hline
$m_z$ & $183\,{\rm GeV}$ & $5.28\times10^{-3}$
\\ \hline
$m_z$ & $500\,{\rm GeV}$ & $7.95\times10^{-5}$
\\ \hline
$2m_Z$ & $183\,{\rm GeV}$ & $2.87\times10^{-3}$
\\ \hline
$2m_Z$ & $500\,{\rm GeV}$ & $1.82\times10^{-4}$
\\ \hline
$900\,{\rm GeV}$ & $183\,{\rm GeV}$ & $2.51\times10^{-3}$
\\ \hline
$900\,{\rm GeV}$ & $500\,{\rm GeV}$ & $2.39\times10^{-4}$
\\ \hline
$1200\,{\rm GeV}$ & $183\,{\rm GeV}$ & $2.50\times10^{-3}$
\\ \hline
$1200\,{\rm GeV}$ & $500\,{\rm GeV}$ & $2.35\times10^{-4}$
\\ \hline
$1400\,{\rm GeV}$ & $183\,{\rm GeV}$ & $2.50\times10^{-3}$
\\ \hline
$1400\,{\rm GeV}$ & $500\,{\rm GeV}$ & $2.33\times10^{-4}$
\\ \hline
\end{tabular}
\caption{\label{f5fixedmh} Values of the $|f_5|$ contribution for the curves shown in the lower graph of Fig.~\ref{2Df5graphs}, at $\sqrt{s}=183\,{\rm GeV}$ and $\sqrt{s}=500\,{\rm GeV}$.}
\end{table}
displays $|f_5|$ contributions, for the variety of considered values of the heavy-neutrino mass, near the threshold (we use $\sqrt{s}=183\,{\rm GeV}$) and at the reference value $\sqrt{s}=500\,{\rm GeV}$, for the CME, as well. The largest $|f_5|$ contributions, for all $m_h$, correspond to CMEs next to $\sqrt{s}=2m_Z$ threshold. For CMEs $\sqrt{s}\gtrsim500\,{\rm GeV}$, on the other hand, the curves corresponding to $m_h=900\,{\rm GeV}, 1200\,{\rm GeV}, 1400\,{\rm GeV}$ seem to dominate within the $\sqrt{s}$-range considered for the graph. Note that, as we pointed out in the previous paragraph, the largest contributions at this CME are smaller, by about one order of magnitude, than projected ILC sensitivity to $f_5$, as estimated in Ref.~\cite{RaSi}.
\\

\section{Summary and conclusions}
\label{concs}
Since the measurement of neutrino oscillations, which incarnates sound evidence supporting massiveness of neutrinos, the mechanism behind neutrino-mass generation has become a priority in the agenda of theoretical and experimental research. The seesaw mechanism and its variants are means to define massive neutrinos, so the exploration of their phenomenology bears great relevance. Furthermore, the genuine neutrino-mass mechanism is linked to the nature of these particles, which, being both electrically neutral and massive, are described by either Dirac or Majorana fields. The present investigation has been developed within the framework of a seesaw variant in which light neutrinos remain massless at the tree level, while getting their masses radiatively, which enables avoiding huge heavy-neutrino masses, thus opening the possibility of measuring new-physics effects within the reach of sensitivity of future or, perhaps, even current experimental facilities. In the context defined by the neutrino model under consideration, the masses of the heavy neutrinos are restricted to be quasi-degenerate to ensure tininess of light-neutrino masses, which nowadays abide by the stringent constraint $m_{\nu_k}\lesssim0.8\,{\rm eV}$.
\\

The neutrino model considered for this work comes along with $Zn_jn_k$ couplings, of the Standard-Model $Z$ boson with mass-eigenspinor neutrinos $n_j$ and $n_k$, which can be light or heavy. Therefore, one-loop contributions to the vertex $ZZZ$, characterized by virtual-neutrino triangle diagrams, exist, which we addressed in the present paper. While absent at the tree level, the $ZZZ$ coupling is generated at the loop level as long as at least one of the external $Z$-boson fields is assumed to be off the mass shell. Conversely, if the three external $Z$ bosons are taken on shell, this coupling is rendered zero, which is a consequence of Bose symmetry. With this in mind, we explored the one-loop Majorana-neutrino contributions to $ZZZ^*$, where $Z^*$ denotes an off-shell virtual $Z$ boson. This vertex is assumed to be a part of the $Z$-pair production process $e^+e^-\to ZZ$, which takes place in machines such as the currently inoperative Large Electron Positron collider and the future International Linear Collider. The general parametrization of the vertex function for the $ZZZ^*$ coupling comprises two form factors, namely, the factor $f_4$, linked to CP violation, and the $f_5$ factor, which preserves CP symmetry. On the grounds of their superficial degree of divergence, the contributing diagrams were expected to generate ultraviolet divergences, which called for usage of a regularization method. To this aim, we followed the dimensional regularization approach to deal with the calculation, finding both the CP-even and the CP-odd contributions, $f_4$ and $f_5$, from each of the involved Feynman diagrams to be ultraviolet finite and renormalization-scale independent. An aspect of the calculation, worth of comment, is that the Majorana nature of the neutrinos increased the number of contributing Feynman diagrams, in comparison with those diagrams to be considered in the case of Dirac neutrinos, by a factor of 4. 
\\

The last part of the paper was devoted to estimate the resulting contributions to the triple gauge coupling $ZZZ^*$ and then discuss them. Our analytic results for the contributions $f_4$ and $f_5$ are functions on neutrino masses, on the $Z$-boson mass, and on the center-of-mass energy $\sqrt{p^2}=\sqrt{s}$, with $p$ the momentum of the off-shell $Z$ boson. These contributions also bear dependence on the $3\times3$ complex matrix $m_{\rm D}m_{\rm M}^{-1}$, emerged from the neutrino-mass mechanism. In the case of the CP-nonpreserving contributions, given by the factor $f_4$, they were found to emerge as long as the matrix $m_{\rm D}m_{\rm M}^{-1}$ is complex. Otherwise, the contribution vanishes. Note that the Standard Model does not produce such CP-violating effects. For heavy-neutrino masses, $m_h$, within $10\,{\rm GeV}\leqslant m_h\leqslant1500\,{\rm GeV}$ and center-of-mass energies $\sqrt{s}$ ranging from $10\,{\rm GeV}$ to $1500\,{\rm GeV}$, the contributions to the modulus $|f_4|$ were estimated. Taking into account that the center of mass energy is restricted to be larger than $2m_Z$, the threshold below which $Z$-pair production is forbidden, and implementing restrictions by the CMS Collaboration on heavy-neutrino mass, we find that contributions might be as large as $\sim 10^{-7}$, which is 3 orders of magnitude below the current best constraint, by the CMS Collaboration. Regarding the CP-even contribution, characterized by the factor $f_5$, in the general parametrization of the vertex function $ZZZ^*$, it exists no matter whether the matrix $m_{\rm D}m_{\rm M}^{-1}$ si complex. Again, heavy-neutrino masses running within $10\,{\rm GeV}\leqslant m_h\leqslant1500\,{\rm GeV}$ and CMEs $\sqrt{s}$ running from $10\,{\rm GeV}$ to $1500\,{\rm GeV}$ were considered. For reference, the one-loop Standard-Model contribution, which is CP conserving, has been reported to vary, with the center of mass energy, from $\sim10^{-4}$ to $\sim10^{-3}$. CP-even contributions from Majorana neutrinos to $|f_5|$ were found to be larger than those which violate CP symmetry. We estimated $|f_5|$ contributions as large as $\sim10^{-4}$, at $\sqrt{s}=500\,{\rm GeV}$, which is comparable to current bounds by the CMS Collaboration. It is also one order of magnitude below the Standard-Model Contribution at $\sqrt{s}=500\,{\rm GeV}$ and one order of magnitude smaller than projected $f_5$-sensitivity of the International Linear Collider at $\sqrt{s}=500\,{\rm GeV}$.

\section*{Acknowledgements}
\noindent
We acknowledge financial support from Conahcyt (M\'exico). M.S. Acknowledges funding by Conahcyt, though the program ``Estancias Posdoctorales por M\'exico 2022''.

\appendix*

\section{The coefficients $\xi_n$}
\label{xinappendix}
The definitions of the $\xi_n$'s, utilized to write down Eq.~(\ref{f5explicit}), are given below:
\begin{eqnarray}
&&
\xi_1=-2\hat\rho^4m_Z\sqrt{m_Z^2-4 m_h^2} \Big(\big(4 \left(\hat\rho ^2-1\right)^2 m_Z^4
\nonumber \\ &&
+4s \left(\hat\rho ^2+1\right)^2 m_Z^2-2 s^2 \left(\hat\rho^4+1\right)\big) m_h^2
\nonumber \\ &&
+2 m_hm_\ell \left(4m_Z^2-s\right) s \left(\hat\rho ^2-1\right)^2 \cos2\phi
\nonumber \\ &&
-4 m_\ell^2 m_Z^2 \left(m_Z^2-s\right)\left(\hat\rho ^2-1\right)^2
\nonumber \\ &&
-m_Z^2 s \left(2m_Z^2+s\right) \left(\hat\rho ^4-\hat\rho ^2+1\right)\Big),
\end{eqnarray}
\begin{eqnarray}
&&
\xi_2=-2 \hat\rho^4m_Z^2 \sqrt{s\left(s-4 m_h^2\right)} \Big(-2 \big(m_Z^2 \left(3\hat\rho ^4+2 \hat\rho ^2+3\right)
\nonumber \\ &&
-2 s \hat\rho^2\big) m_h^2-2m_hm_\ell \left(4 m_Z^2-s\right) \left(\hat\rho ^2-1\right)^2\cos2\phi
\nonumber \\ &&
-2 m_\ell^2\left(m_Z^2-s\right) \left(\hat\rho ^2-1\right)^2
\nonumber \\ &&
+m_Z^2\left(2 m_Z^2+s\right)\left(\hat\rho ^4-\hat\rho^2+1\right)\Big),
\end{eqnarray}
\begin{eqnarray}
&&
\xi_3=-\hat\rho ^2\left(\hat\rho ^2-1\right)^2(s-m_Z^2)(m_\ell^2-m_h^2)
\nonumber \\ &&
\times
\Big(\big(2\left(10 \hat\rho ^2-9\right) m_Z^2+s \left(3-2 \hat\rho^2\right)\big) m_\ell^2
\nonumber \\ &&
+4 m_hm_\ell \left(s-4m_Z^2\right) \cos2\phi+2 m_h^2\left(s-10 m_Z^2\right) \hat\rho^2
\nonumber \\ &&
+\left(m_h^2+4m_Z^2\right) \left(2 m_Z^2+s\right)\Big),
\end{eqnarray}
\begin{eqnarray}
&&
\xi_4=-\sqrt{(m_\ell^2-(m_h+m_Z)^2)(m_\ell^2-(m_h-m_Z)^2)}
\nonumber \\ &&
\times
2\hat\rho^2\left(\hat\rho ^2-1\right)^2\Big(-4 m_\ell^2 m_Z^4+4 m_h^2 m_Z^4-4 sm_Z^4
\nonumber \\ &&
-2 s^2 m_Z^2+16 m_\ell^2 s m_Z^2-3m_\ell^2 s^2-m_h^2 s^2
\nonumber \\ &&
+2 (m_\ell^2-m_h^2)\left(4 m_Z^4-8 sm_Z^2+s^2\right) \hat\rho ^2
\nonumber \\ &&
+4 m_\ell m_h \left(4m_Z^2-s\right) s \cos2\phi\Big),
\end{eqnarray}
\begin{eqnarray}
&&
\xi_5=2m_Z^2\sqrt{m_\ell^4-2\left(m_h^2+s\right)m_\ell^2+\left(m_h^2-s\right)^2} 
\nonumber \\ &&
\times
\hat\rho^2 \left(\hat\rho^2-1\right)^2\Big(\left(2 \left(7-6 \hat\rho ^2\right)m_Z^2+s \left(6 \hat\rho ^2-5\right)\right) m_\ell^2
\nonumber \\ &&
+4m_hm_\ell \left(4 m_Z^2-s\right) \cos (2 \phi )+6 m_h^2 \left(2 m_Z^2-s\right) \hat\rho^2
\nonumber \\ &&
+\left(m_h^2-2 m_Z^2\right) \left(2m_Z^2+s\right)\Big),
\end{eqnarray}
\begin{eqnarray}
&&
\xi_6=2\left(\hat\rho^2-1\right)^3m_Z\sqrt{m_Z^2-4 m_\ell^2}\Big(-2 \big(\left(2 m_h^2+s\right) \hat\rho^2
\nonumber \\ &&
+s\big) m_Z^4-s \left(\left(s-4 m_h^2\right) \hat\rho^2+s\right) m_Z^2+2 m_\ell^2 \big(2 \hat\rho ^2m_Z^4
\nonumber \\ &&
+2 s \left(\hat\rho ^2+4\right) m_Z^2-s^2 \left(\hat\rho^2+2\right)\big)
\nonumber \\ &&
+2 m_\ell m_h \left(4m_Z^2-s\right) s \hat\rho ^2 \cos2\phi\Big),
\end{eqnarray}
\begin{eqnarray}
&&
\xi_7=2\left(\hat\rho ^2-1\right)^3 m_Z^2 \sqrt{s\left(s-4 m_\ell^2\right)}\Big(2\left(\hat\rho ^2+1\right) m_Z^4
\nonumber \\ &&
+\left(\left(s-2m_h^2\right) \hat\rho ^2+s\right) m_Z^2+2 m_h^2 s\hat\rho ^2-2 m_\ell^2 \big(m_Z^2 \big(3 \hat\rho^2
\nonumber \\ &&
+8\big)-2 s\big)+2 m_\ell m_h \left(s-4m_Z^2\right)\hat\rho^2 \cos2\phi\Big),
\end{eqnarray}
\begin{equation}
\xi_8=4 \left(\hat\rho ^2-1\right)^3m_Z^2s(s-m_Z^2)\left(m_Z^4+m_\ell^2 \left(s-4 m_Z^2\right)\right),
\end{equation}
\begin{equation}
\xi_9=-4 \hat\rho ^6m_Z^2 s(s-m_Z^2)\left(m_Z^4+m_h^2 \left(s-4m_Z^2\right)\right),
\end{equation}
\begin{eqnarray}
&&
\xi_{10}=-2\hat\rho^4m_Z^2s \left(\hat\rho^2-1\right)^2 \Big(2 \left(s-m_Z^2\right)m_\ell^4
\nonumber \\ &&
+\left(4 (m^2_Z-m^2_h)m_Z^2+s^2-2 \left(m_h^2+m_Z^2\right) s\right)m_\ell^2
\nonumber \\ &&
-m_hm_\ell \left(4 m_Z^2-s\right) \left(2m_\ell^2-2 \left(m_h^2+m_Z^2\right)+s\right) \cos2\phi
\nonumber \\ &&
+2m_Z^2(m^2_h-m^2_Z) \left(3m_h^2+m_Z^2-s\right)\Big),
\end{eqnarray}
\begin{eqnarray}
&&
\xi_{11}=-2\hat\rho^4\left(\hat\rho^2-1\right)^2m_Z^2\Big(-4 \left(m_\ell^2-m_h^2\right)^2m_Z^4
\nonumber \\ &&
-4 \big(-m_\ell^4+m_h^4+m_Z^4-4m_h^2 m_Z^2\big) s m_Z^2+3 m_h^2s^3
\nonumber \\ &&
+\left(2 m_h^4-13 m_Z^2 m_h^2+4m_Z^4-m_\ell^2 \left(2 m_h^2+3m_Z^2\right)\right) s^2
\nonumber \\ &&
-m_\ell m_h \left(2m_\ell^2-2 m_h^2-s\right) s \left(s-4 m_Z^2\right)\cos2\phi\Big),
\end{eqnarray}
\begin{eqnarray}
&&
\xi_{12}=-2 \hat\rho^2\left(\hat\rho^2-1\right)^3m_Z^2s  \Big(-6 m_Z^2 m_\ell^4+2\big(m_h^2
\nonumber \\ &&
+m_Z^2\big) \left(2 m_Z^2+s\right)m_\ell^2-m_h \left(4 m_Z^2-s\right) \big(2\big(m_\ell^2-m_h^2
\nonumber \\ &&
+m_Z^2\big)-s\big) \cos2\phi m_\ell+2 \left(m_Z^3-m_h^2m_Z\right)^2-m_h^2 s^2
\nonumber \\ &&
-2\left(m_h^4-m_Z^2 m_h^2+m_Z^4\right)s\Big),
\end{eqnarray}
\begin{eqnarray}
&&
\xi_{13}=-2 \hat\rho ^2\left(\hat\rho^2-1\right)^3m_Z^2\Big(4 \left(m_\ell^2-m_h^2\right)^2m_Z^4+4 \big(m_\ell^4
\nonumber \\ &&
-4 m_Z^2m_\ell^2-m_h^4+m_Z^4\big) s m_Z^2-3m_\ell^2 s^3+\big(-2 m_\ell^4
\nonumber \\ &&
+\left(2 m_h^2+13m_Z^2\right) m_\ell^2-4 m_Z^4+3m_h^2m_Z^2\big) s^2
\nonumber \\ &&
+m_\ell m_h \left(4m_Z^2-s\right) s \left(2 m_\ell^2-2 m_h^2+s\right)\cos2\phi\Big),
\end{eqnarray}
\begin{eqnarray}
&&
\xi_{14}=-2m_Z^4s \left(3 \hat\rho ^6-3\hat\rho ^4+1\right)(s-4m_Z^2).
\end{eqnarray}

\end{document}